\documentclass[journal]{IEEEtran}

\usepackage{graphicx}
\usepackage{subcaption}
\usepackage{cite}
\usepackage{amsmath}
\usepackage{amssymb}
\usepackage{bm}
\usepackage{amsfonts}
\usepackage{array}
\usepackage{url}
\usepackage{stfloats}
\usepackage{enumitem} 
\usepackage{xcolor,soul,framed}
\usepackage{mdwmath}
\usepackage{mdwtab}
\usepackage{lineno,hyperref}
\usepackage{bbm}
\usepackage{algorithm}
\usepackage{algpseudocode}
\usepackage{booktabs}
\usepackage{multirow}

\begin{document}

\title{Token Communication for Multimodal Large Language Model}

% author
\author{Jingkai~Ying,~\IEEEmembership{Graduate Student Member,~IEEE,}
        Zhijin~Qin,~\IEEEmembership{Senior Member,~IEEE,} \\
        Yuan~Shen,~\IEEEmembership{Senior Member,~IEEE,}
        and~Khaled~B.~Letaief,~\IEEEmembership{Fellow,~IEEE}% <-this % stops a space
\thanks{Jingkai Ying, Zhijin Qin and Yuan Shen are with the Department of Electronic Engineering, Tsinghua University, Beijing 100084, China, and also with the State Key Laboratory of Space Network and Communications, Beijing 100084, China (e-mail: yjk23@mails.tsinghua.edu.cn; qinzhijin@tsinghua.edu.cn; shenyuan\_ee@tsinghua.edu.cn).}% <-this % stops a space
\thanks{Khaled B. Letaief is with the Department of Electronic
and Computer Engineering, The Hong Kong University of Science and
Technology, Hong Kong (e-mail: eekhaled@ust.hk).}% <-this % stops a space
}

% paper headers

\maketitle

% abstract
\begin{abstract}
With the broad success of the Transformer architecture, token is becoming a new basic information processing unit.
This trend is especially evident in multimodal large language models (MLLMs), where both visual and textual information are represented and processed as tokens.
With the rapid deployment of MLLMs, the efficient transmission of tokens has become increasingly important.
This paper investigates how to reduce the amount of transmitted data during interactions with MLLMs while preserving their multimodal understanding performance.
To address this problem, we propose a token communication framework tailored to MLLMs.
In the proposed framework, a neural codec is integrated into the vision tokenizer to control the number of transmitted bits.
At the receiver, the decoded latents are processed through two paths.
The decoder reconstructs image as a reconstruction prior, while the adapter converts latents into visual tokens and injects them into an intermediate layer of the vision tokenizer.
To make the injected tokens suitable for MLLMs, we further design a two-stage visual-language semantic alignment training scheme.
The adapter is first warmed up by a distillation loss and then aligned with textual semantics through an alignment loss.
An adaptive adapter is also introduced through feature-wise linear modulation, allowing one adapter to support multiple codec rates.
Extensive simulations on various MLLM benchmarks show that, under the same amount of transmitted data, the proposed scheme achieves better task performance than other image processing schemes for MLLMs.
\end{abstract}

% keywords

\section{Introduction}
\label{sec:Introduction}

Since the release of ChatGPT in late 2022, large language models (LLMs) have achieved remarkable progress \cite{achiam2023gpt, guo2025deepseek}.
These models have demonstrated strong capabilities in logical reasoning and language understanding, achieving strong performance in tasks such as translation, writing, and coding.
The success of LLMs has further been extended to the multimodal field.
Multimodal large language models (MLLMs) enable models to process visual information and perform multimodal understanding tasks \cite{team2023gemini, bai2025qwen3}.
Both LLMs and MLLMs are built upon the Transformer \cite{vaswani2017attention} architecture and therefore use tokens as their basic information processing units.
With the growing applications of these large artificial intelligence (AI) models, the daily token consumption has reached the trillion scale \cite{aubakirova2026state}.
This trend further highlights the increasing importance of tokens.

\begin{figure}[!t]
    \centering
    \includegraphics[width=0.4\textwidth]{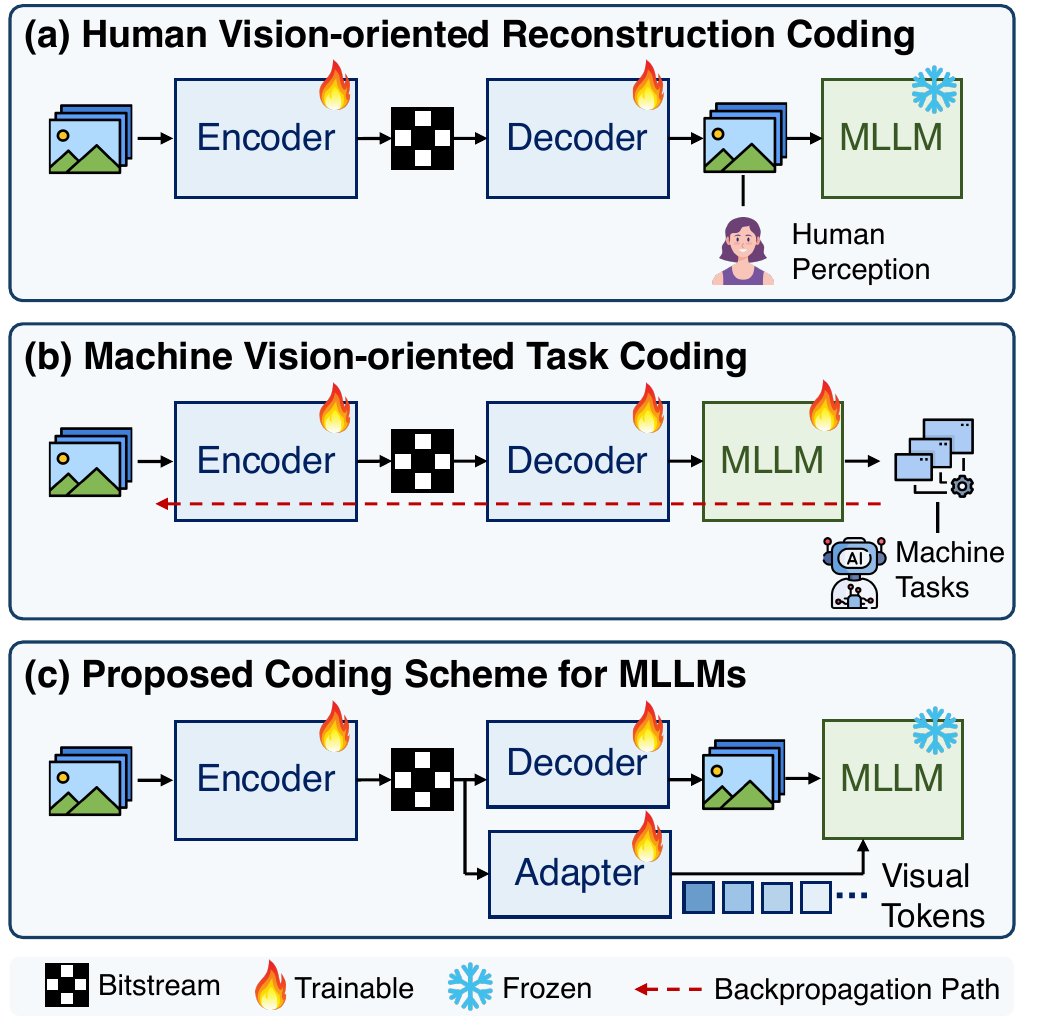}
    \caption{Overview of frameworks when different coding schemes are applied to MLLMs.}
    \label{fig:coding_schemes}
\end{figure}

Focusing on token, a new type of information unit, researchers in the communication society have investigated the topic of token communication \cite{qiao2025token, devoto2026adaptive, liu2025resitok, ying2026joint}.
These pioneering works mainly study how to exploit token representations with rich semantics and powerful pre-trained models to achieve effective and reliable communication.
They focus on tasks such as classification \cite{devoto2026adaptive} and reconstruction \cite{qiao2025token, liu2025resitok, ying2026joint}.
Experimental results have shown the advantages of token communication under low-rate and severe channel conditions.
However, for large AI models, which are the major consumers of tokens, targeted studies on token communication for corresponding tasks remain limited.

In this paper, we design a token communication scheme for MLLMs.
The goal is to reduce the transmitted data volume required for interacting with MLLMs while preserving their performance on diverse tasks.
Image compression is a major approach for reducing the number of transmitted bits.
Existing image compression methods can mainly be divided into human vision-oriented reconstruction coding and machine vision-oriented task coding.
Directly applying these two types of methods to MLLM-oriented compression leads to the schemes shown in Fig. \ref{fig:coding_schemes}(a) and Fig. \ref{fig:coding_schemes}(b).
For the human vision-oriented scheme, it follows the existing paradigm for interacting with MLLMs.
Images captured by edge devices are encoded into compact bitstreams and transmitted to the cloud server.
The decoder at the server reconstructs images for human perception \cite{jia2025towards, he2022elic}.
The reconstructed images are then fed into the MLLM.
This scheme optimizes the representation for human vision.
However, a human-perception-oriented representation is not necessarily optimal for MLLMs \cite{fu2026cache, xiao2025transmission}.
The machine-vision-oriented scheme, commonly called coding for machines \cite{yang2024video}, can achieve more compact compression for specific machine tasks, such as classification, object detection, and semantic segmentation.
However, this scheme usually requires task-specific end-to-end training, which is difficult to satisfy when MLLMs are used for general multimodal understanding tasks.
Therefore, we propose a coding scheme tailored for MLLMs, as shown in Fig. \ref{fig:coding_schemes}(c).
Inspired by existing studies on token communication, our scheme fully exploits powerful vision tokenizers to provide more native token representations for MLLMs.
By incorporating the encoder for compression into the vision tokenizer, our approach transmits intermediate visual tokens without compromising compression efficiency.

Specifically, we propose a token-based coding scheme for MLLMs.
To control the number of bits transmitted through the channel, we employ an entropy model-based neural codec for bit compression.
To provide informative token representations for MLLMs, the receiver uses both a decoder and an adapter to transform compression-oriented latents into pixel-domain images and feature-domain visual tokens.
The compressed image serves as a reconstruction prior for subsequent visual tokens.
The visual tokens converted by the adapter are injected into an intermediate layer of the vision tokenizer.
This design enables the MLLM to receive suitable visual token inputs.
In this information processing flow, dedicated design of the adapter is required.
Because the visual tokens injected by the adapter should be semantically aligned with the corresponding textual descriptions as much as possible.
For visual tokenizers based on visual-language alignment in MLLMs \cite{radford2021learning, zhai2023sigmoid, tschannen2025siglip}, the combination of the reconstruction prior and the adapter output is meaningful only under such alignment.
Otherwise, the injected adapter output can degrade the quality of the final visual tokens fed into the MLLM.
Therefore, we design a visual-language semantic alignment training scheme.
Through a two-stage training algorithm with distillation-based initialization and Sigmoid Language-Image Pre-training (SigLIP) loss \cite{zhai2023sigmoid}, learnable parameters of the adapter can be optimized.
This training scheme adapts outputs of the adapter for MLLMs.
In addition, we design the adapter to be adaptive to the quantization parameter (QP) of the codec.

The main contributions of this paper can be summarized as follows:
\begin{itemize}
\item A token communication framework is developed to provide informative visual tokens for MLLMs.
In this framework, the neural codec is integrated into the vision tokenizer to control the number of bits required for token transmission. 
At the receiver, a decoder provides the reconstruction prior, while an adapter converts compression-oriented features into visual tokens and injects them into the vision tokenizer.
\item To improve the effectiveness of visual token injection by adapter, a visual-language semantic alignment training scheme is proposed.
Based on effective extraction of visual and textual semantic features, the adapter is first warmed up with a distillation loss and then semantically aligned with the SigLIP loss.
\item A QP-adaptive adapter is introduced so that neural codecs with different QPs can share the same adapter. To this end, feature-wise linear modulation (FiLM) \cite{perez2018film} is utilized in the adapter to incorporate QP.
By randomly sampling QP during training, the adapter can transform latents under various QPs into suitable visual tokens.
\end{itemize}

The remainder of this paper is organized as follows.
Section \ref{sec:Related_Work} reviews the related work.
Section \ref{sec:Overview_of_the_Proposed_Token_Communication_framework_for_MLLMs} presents the framework of the proposed token communication for MLLMs.
Section \ref{sec:Model_Design_and_Training_Scheme} details the approach for aligning compression-oriented visual features with textual semantics.
Section \ref{sec:Simulation_Results} provides the simulation results.
Section \ref{sec:Conclusion} concludes the paper.

\textit{Notations}: For a set $\mathcal{X}$, $|\mathcal{X}|$ denotes its cardinality.
For a vector, matrix, or tensor $\boldsymbol{x}$, $\|\boldsymbol{x}\|$ denotes the Euclidean norm after vectorization.
For a complex-valued scalar $z$, $|z|$ and $z^{*}$ denote its modulus and complex conjugate, respectively.
The operators $(\cdot)^{T}$ and $(\cdot)^{H}$ denote transpose and Hermitian transpose, respectively.
The expectation operator is denoted by $\mathbb{E}[\cdot]$.
The Hadamard product is denoted by $\odot$.
The indicator function is denoted by $\mathbbm{1}(\cdot)$.
The notation $\mathcal{CN}(\boldsymbol{\mu},\boldsymbol{\Sigma})$ denotes a complex Gaussian distribution with mean vector $\boldsymbol{\mu}$ and covariance matrix $\boldsymbol{\Sigma}$.
The spaces of real-valued and complex-valued arrays with dimensions $d_1 \times \cdots \times d_K$ are denoted by $\mathbb{R}^{d_1 \times \cdots \times d_K}$ and $\mathbb{C}^{d_1 \times \cdots \times d_K}$, respectively.

\section{Related Work}
\label{sec:Related_Work}

In this section, we review the related work on token communication and image neural codec.

\subsection{Token Communication}
Motivated by the strong performance of Transformers across different modalities, pioneering works on token communication focused on using powerful tokenizers to obtain compact and semantically rich token representations \cite{qiao2025token}.
Based on these token representations, Transformer-based models can also be used at the receiver to predict lost tokens and support cross-modal auxiliary reconstruction.
These capabilities demonstrate the superiority of token communication.
To further improve the robustness of token indices transmitted over channels, some works have investigated unequal error protection (UEP) \cite{liu2025resitok, men2026video, zhang2026tokencom, lee2025semantic}.
For tokens with different levels of importance, these works adjust the quantization precision, select suitable modulation and coding scheme (MCS) \cite{liu2025resitok, men2026video}, or apply UEP channel coding \cite{zhang2026tokencom}.
Other works have studied how to characterize token importance for specific tasks and pack tokens accordingly\cite{lee2025semantic}.

The tokenizers used in the above works are based on vector quantization (VQ), providing discrete tokens.
The indices of these tokens naturally provide compact bit representations for digital communication.
This property is consistent with VQ-based semantic communication systems \cite{fu2023vector}, while token communication usually relies on more powerful pre-trained tokenizers.
As a core technique in semantic communication, joint semantic-channel coding (JSCC) has also been applied to token communication \cite{devoto2026adaptive, xiao2025transmission, jiang2026tokencom}.
Since JSCC can directly map the tokenizer outputs to channel symbols, these works can adopt continuous tokenizers.
For MLLMs, continuous tokenizers \cite{zhai2023sigmoid, radford2021learning, tschannen2025siglip} are necessary to achieve stronger multimodal understanding performance \cite{wu2025vila}.
Two very recent works \cite{jiang2026tokencom,xiao2025transmission} have investigated token communication for MLLMs.
These works focus on robust joint semantic-channel coding (JSCC) design and require end-to-end training involving the LLM.
By contrast, this paper focuses on reducing the number of transmitted bits and designs a lightweight adapter-based training scheme with semantic alignment.

\begin{figure*}[!t]
    \centering
    \includegraphics[width=0.85\textwidth]{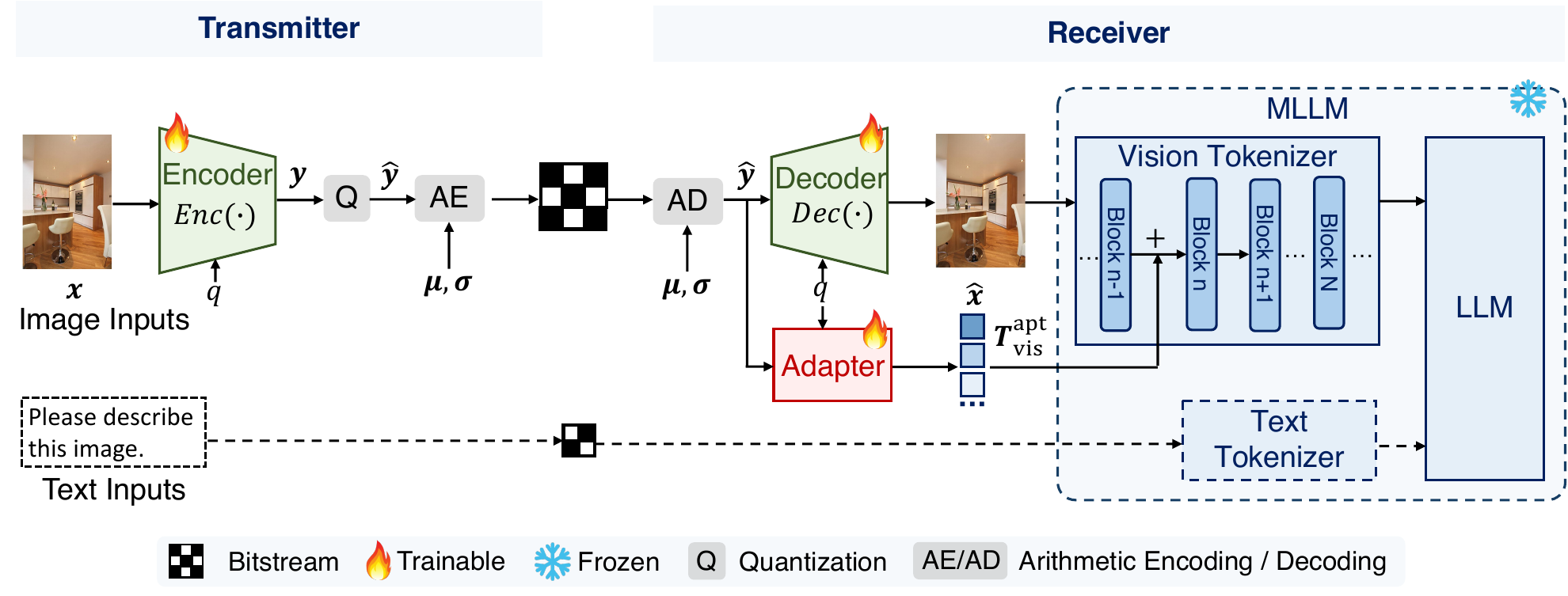}
    \caption{The proposed token communication framework for MLLMs. After compression at the transmitter, the image is processed by the decoder and the adapter at the receiver, and the resulting outputs are fed into the MLLM.}
    \label{fig:framework}
\end{figure*}

\subsection{Image Neural Codec}
For human vision-oriented reconstruction, image neural codecs replace the manually designed modules in traditional image codecs with neural networks \cite{jia2025towards, he2022elic}.
With powerful nonlinear transforms and entropy models, they have outperformed traditional codecs such as BPG \cite{bellard2015bpg} and VTM \cite{vtm}.
Current research on neural codecs is gradually shifting from pixel-level optimization to human-perception optimization at ultra-low bitrate (\textit{e.g.} $<$ 0.05 bits per pixel) \cite{jia2024generative, cao2026progic}.
Another class of image neural codecs is designed for machine vision \cite{yang2024video}.
Since these codecs are trained for specific tasks, they can achieve more aggressive compression \cite{chen2023transtic}.

However, human-perception-oriented representations are not necessarily optimal for MLLMs \cite{fu2026cache, xiao2025transmission}.
As for machine-vision-oriented coding methods, they require task-specific end-to-end training \cite{yang2024video}.
This requirement is clearly unsuitable for MLLMs, which contain billions of parameters and are expected to perform diverse multimodal understanding tasks.
Studies on neural codecs for MLLMs aim to obtain visual inputs that are more suitable for MLLMs from compressed latents.
A token extractor and a pre-editing network have been introduced to reduce the impact of compression-induced distortion on MLLMs \cite{li2025high}.
To introduce fewer modifications to the codec, adapter-based schemes based on semantic alignment \cite{kao2025bridging} and multi-level distillation \cite{liu2026when} have also been proposed.

\section{Overview of the Proposed Token Communication framework for MLLMs}
\label{sec:Overview_of_the_Proposed_Token_Communication_framework_for_MLLMs}

In this section, we introduce the proposed token communication framework, including the entire information processing flow, the function of each model and their corresponding inputs and outputs. In addition, we present the focused problem of this framework.

\subsection{Formulation of System Models}
The overall illustration of the proposed framework is provided in Fig. \ref{fig:framework}.
During the interaction between an edge device and a cloud-based MLLM, the edge device sends both images and text to the MLLM.
For image transmission, the encoder at the transmitter converts the pixel-domain image into compression-oriented latents.
To reduce the number of transmitted bits through entropy coding, the continuous latents are first quantized.
At the receiver, the quantized latents are losslessly reconstructed after arithmetic decoding.
The reconstructed latents are then processed by the decoder to recover a pixel-domain image and by the adapter to produce vision tokens suitable for MLLM input.
The reconstructed image is fed into the vision tokenizer from the beginning.
After patchification and embedding projection,, it is converted into visual tokens.
The vision tokens produced by the adapter are injected into the intermediate Transformer blocks of the vision tokenizer.
They are fused with the visual tokens produced from the reconstructed image through addition.
For text transmission, the text is also encoded and decoded.
At the receiver, the decoded text is tokenized and embedded before being fed into the large model.
Since this paper focuses on image transmission for MLLMs, the text transmission part is simplified in Fig. \ref{fig:framework}.
In the remainder of this paper, we only discuss the image processing part.

Specifically, let $\boldsymbol{x}\in\mathbb{R}^{C\times H\times W}$ denote the input image, where $C$ denotes the number of color channels, and $H$ and $W$ denote the image height and width, respectively.
At the transmitter, a deep learning (DL)-based encoder $\operatorname{Enc}(\cdot)$ with learnable parameters $\boldsymbol{\theta}_{\operatorname{enc}}$ performs a nonlinear analysis transform as
\begin{align}
\boldsymbol{y}
=
\operatorname{Enc}\left(\boldsymbol{x}; q,\boldsymbol{\theta}_{\operatorname{enc}}\right),
\end{align}
where $\boldsymbol{y}\in\mathbb{R}^{C'\times H'\times W'}$ denotes the compression-oriented latent representation.
Here, $q\in\mathcal{Q}=\{0,1,2,\cdots\}$ denotes the quantization parameter (QP) that controls the compressed bitrate.
The proposed framework can support different types of neural codecs.
For a rate-adaptive neural codec \cite{jia2025towards}, $q$ corresponds to the index of a learnable channel-wise scaling vector.
For a fixed rate neural codec \cite{he2022elic}, $q$ indicates the index of the model weights trained for a specific rate point.

Subsequently, the latent representation $\boldsymbol{y}$ is quantized and then entropy-coded into a bitstream for efficient transmission.
At the receiver, the received bitstream is decoded to recover the quantized latent representation $\hat{\boldsymbol{y}}$.
This process can be formulated as
\begin{align}
\hat{\boldsymbol{y}}
=
\operatorname{AD}\left(
\operatorname{AE}\left(
\operatorname{Q}\left(\boldsymbol{y}\right);
\boldsymbol{\mu},\boldsymbol{\sigma}
\right);
\boldsymbol{\mu},\boldsymbol{\sigma}
\right),
\end{align}
where $\operatorname{Q}(\cdot)$ denotes quantization, and $\operatorname{AE}(\cdot)$ and $\operatorname{AD}(\cdot)$ denote arithmetic encoding and arithmetic decoding, respectively.
The variables $\boldsymbol{\mu}$ and $\boldsymbol{\sigma}$ are the Gaussian probability model parameters of $\hat{\boldsymbol{y}}$ estimated by the entropy model, which can be expressed as
\begin{align}
(\boldsymbol{\mu},\boldsymbol{\sigma})
=
\Phi\left(\boldsymbol{y};\boldsymbol{\theta}_{\operatorname{entropy}}\right),
\end{align}
where $\Phi(\cdot)$ denotes the entropy model with learnable parameters $\boldsymbol{\theta}_{\operatorname{entropy}}$.
More details about the above hyperprior-based entropy estimation in neural codecs can be found in \cite{balle2018variational}.

For the reconstructed latent representation $\hat{\boldsymbol{y}}$, two information paths are used at the receiver.
In the first path, a decoder $\operatorname{Dec}(\cdot)$ performs a nonlinear synthesis transform to reconstruct the compressed image $\hat{\boldsymbol{x}}$ in the pixel domain as
\begin{align}
\hat{\boldsymbol{x}}
=
\operatorname{Dec}\left(\hat{\boldsymbol{y}}; q,\boldsymbol{\theta}_{\operatorname{dec}}\right),
\end{align}
where $\boldsymbol{\theta}_{\operatorname{dec}}$ denotes the learnable parameters of the decoder and $q$ is the same as in the encoding process.
This path provides pixel-level information as a strong reconstruction prior for the MLLM input.
Such a prior helps the MLLM handle inputs that require detailed visual content, such as posters and optical character recognition (OCR) images.

In the second path, a Transformer block-based adapter $\operatorname{Apt}(\cdot)$ with learnable parameters $\boldsymbol{\theta}_{\operatorname{apt}}$ converts the compression-oriented latent representation into visual tokens as
\begin{align}
\boldsymbol{T}^{\operatorname{apt}}_{\operatorname{vis}}
=
\operatorname{Apt}\left(\hat{\boldsymbol{y}}; q,\boldsymbol{\theta}_{\operatorname{apt}}\right),
\end{align}
where $\boldsymbol{T}^{\operatorname{apt}}_{\operatorname{vis}}\in\mathbb{R}^{L\times D}$ denotes the visual tokens produced by the adapter.
Here, $L$ is the number of visual tokens, and $D$ is the dimension of token embeddings.

The reconstructed image $\hat{\boldsymbol{x}}$ is fed into the vision tokenizer from the beginning.
After patchification and linear embedding projection, it is converted into reconstructed visual tokens as
\begin{align}
\boldsymbol{T}^{\operatorname{rec},0}_{\operatorname{vis}}
=
\operatorname{Embed}
\left(
\operatorname{Patch}
\left(
\hat{\boldsymbol{x}}
\right);
\boldsymbol{\theta}_{\operatorname{emb}}
\right),
\end{align}
where $\operatorname{Patch}(\cdot)$ denotes the patchification operation, and $\operatorname{Embed}(\cdot)$ denotes the linear embedding projection with parameters $\boldsymbol{\theta}_{\operatorname{emb}}$.
Subsequently, the reconstructed visual tokens are processed by the first $n-1$ Transformer blocks $\mathcal{V}_{i}(\cdot)$ with parameters $\boldsymbol{\theta}_{\mathcal{V}_{i}}$ of the vision tokenizer:
\begin{align}
\boldsymbol{T}^{\operatorname{rec},i}_{\operatorname{vis}}
=
\mathcal{V}_{i}\left(
\boldsymbol{T}^{\operatorname{rec},i-1}_{\operatorname{vis}};
\boldsymbol{\theta}_{\mathcal{V}_{i}}
\right),
\quad
i=1,\cdots,n-1.
\end{align}
The adapter-produced visual tokens are injected into the intermediate layer and fused with the visual tokens produced from the reconstruction prior by addition:
\begin{align}
\boldsymbol{T}_{\operatorname{vis}}
=
\mathcal{V}_{n:N_{\mathcal{V}}}
\left(
\boldsymbol{T}^{\operatorname{rec},n-1}_{\operatorname{vis}}
+
\alpha \boldsymbol{T}^{\operatorname{apt}}_{\operatorname{vis}};
\boldsymbol{\theta}_{\mathcal{V}_{n:N_{\mathcal{V}}}}
\right),
\end{align}
where $\mathcal{V}_{n:N_{\mathcal{V}}}(\cdot)$ denotes the remaining Transformer blocks from the $n$-th block to the final block and $\boldsymbol{\theta}_{\mathcal{V}_{n:N_{\mathcal{V}}}}$ denotes their model parameters.
The scalar $\alpha$ is a hyper-parameter that controls the strength of the adapter-based token injection.
After proper semantic alignment during training, the adapter can provide semantically informative visual inputs for the MLLM.
This design is especially useful in the low-rate region, where the reconstructed image has lost substantial visual information.

Finally, the visual tokens $\boldsymbol{T}_{\operatorname{vis}}$ and the text tokens $\boldsymbol{T}_{\operatorname{text}}$ are concatenated and processed by LLM for autoregressive next-token prediction:
\begin{align}
p\left(\boldsymbol{s}\mid \boldsymbol{T}_{\operatorname{vis}},\boldsymbol{T}_{\operatorname{text}}\right)
=
\prod_{t=1}^{T}
p\left(
s_t
\mid
s_{<t},
\boldsymbol{T}_{\operatorname{vis}},
\boldsymbol{T}_{\operatorname{text}};
\boldsymbol{\theta}_{\operatorname{LLM}}
\right),
\end{align}
where $\boldsymbol{s}=\{s_t\}_{t=1}^{T}$ denotes the output token sequence with length $T$.
Here, $p(\cdot)$ denotes the conditional token distribution.

In summary, the proposed token communication framework effectively combines an image neural codec with the vision tokenizer of an MLLM.
In this framework, the original vision tokenizer, the codec, and the adapter can be viewed as an enlarged vision tokenizer.
The channel separates this enlarged tokenizer into transmitter-side and receiver-side components.
The transmitter outputs intermediate features of the vision tokenizer, while the codec controls the number of transmitted bits.
Different from VQ-based token communication schemes, the proposed framework avoids deploying the entire large vision tokenizer for MLLMs on the edge device.

\subsection{Task Description}
Because MLLMs are trained with original images, the reconstructed images after compression do not necessarily provide the most suitable tokens for MLLMs.
Therefore, the problem considered in this paper is how to obtain visual representations that are more suitable for MLLMs than reconstructed images under a limited number of bits or channel symbols.
These representations are expected to support MLLMs in completing multimodal understanding tasks.
The quality of the visual tokens provided to the MLLM is evaluated by the task performance of the MLLM under the same compression ratio.
In general, the performance achieved by the original image input provides an upper bound for this evaluation.

Under the assumption of perfect transmission, only source compression needs to be considered.
The compression ratio can be measured by bits per pixel (bpp), which is defined as
\begin{align}
\operatorname{bpp}
=
\frac{B}{H\times W},
\end{align}
where $B$ denotes the length of the compressed bitstream, and $H\times W$ denotes the number of pixels in the input image.

When wireless transmission is considered, channel coding and modulation are further required.
Let $R_{\operatorname{c}}=k/n$ denote the channel coding rate, where $k$ is the number of information bits and $n$ is the number of coded bits.
For $M$-ary modulation, each modulation symbol carries $\log_2 M$ bits.
Therefore, the number of channel symbols required for transmitting the compressed bitstream is given by
\begin{align}
N_{\operatorname{ch}}
=
\frac{B}{R_{\operatorname{c}}\times \log_2 M}.
\end{align}
The compression ratio can then be measured by the channel bandwidth ratio (CBR) \cite{bourtsoulatze2019deep}, which is defined as
\begin{align}
\operatorname{CBR}
=
\frac{N_{\operatorname{ch}}}{C\times    H\times W}
=
\frac{\operatorname{bpp}}{C\times R_{\operatorname{c}}\times \log_2 M}.
\end{align}

For wireless transmission, we consider the Rayleigh fading channel.
Let $\boldsymbol{z}\in\mathbb{C}^{N_{\operatorname{ch}}}$ denote the modulated symbol vector after channel coding and modulation.
The received symbols are given by
\begin{align}
\boldsymbol{r}
=
h\cdot\boldsymbol{z}
+
\boldsymbol{n},
\end{align}
where $h$ denotes the channel gain of Rayleigh fading, and $\boldsymbol{n}$ denotes the additive complex Gaussian noise.
Here, the channel gain follows $h\sim\mathcal{CN}(0,1)$, and the noise follows $\boldsymbol{n}\sim\mathcal{CN}(\boldsymbol{0},\sigma_n^2\boldsymbol{I})$. $\sigma_n^2$ denotes the noise power and $\boldsymbol{I}$ denotes the identity matrix.
The received signal power is calculated as
\begin{align}
P_{\operatorname{signal}}
=
\frac{
\mathbb{E}
\left[
\left\|
h\cdot\boldsymbol{z}
\right\|^2
\right]
}
{N_{\operatorname{ch}}}.
\end{align}

Therefore, the signal-to-noise ratio (SNR) in decibels can be defined as
\begin{align}
\operatorname{SNR}
=
10\log_{10}
\frac{P_{\operatorname{signal}}}{\sigma_n^2}.
\end{align}
At the receiver, zero-forcing (ZF) equalization is applied to compensate for the channel fading:
\begin{align}
\hat{\boldsymbol{z}}
=
\frac{h^{*}}{|h|^2}
\cdot
\boldsymbol{r},
\end{align}
where $\hat{\boldsymbol{z}}$ denotes the equalized symbol vector.
The equalized symbols are then demodulated and channel-decoded to recover the received bitstream.

With the above definitions, the considered problem can be expressed as maximizing the MLLM task performance under a rate constraint:
\begin{align}
\max_{\boldsymbol{\Theta}}
\quad
&\mathcal{P}_{\operatorname{MLLM}}
\left(
\boldsymbol{\Theta}
\right), \\
\operatorname{s.t.}
\quad
&\operatorname{bpp}\le R
\quad
\text{or}
\quad
\operatorname{CBR}\le \rho,\ \operatorname{SNR}=\gamma.
\end{align}
$\boldsymbol{\Theta}$ denotes parameters of the neural codec and adapter in proposed framework. $\mathcal{P}_{\operatorname{MLLM}}(\cdot)$ denotes the task performance of the MLLM. $R$, $\rho$ and $\gamma$ are the specific bpp, CBR, and SNR constraints, respectively.

It should be noted that the multimodal understanding tasks refer to tasks where MLLMs receive image-text inputs and generate textual answers according to the text prompt and the image content.
The output of the MLLM is text only.
Image generation tasks, where large AI models output images, are completed by image generation models \cite{wu2025qwen}, or unified MLLMs \cite{wu2025vila}, rather than MLLMs.
Such tasks are beyond the scope of this paper.

\begin{figure*}[!t]
    \centering
    \includegraphics[width=0.85\textwidth]{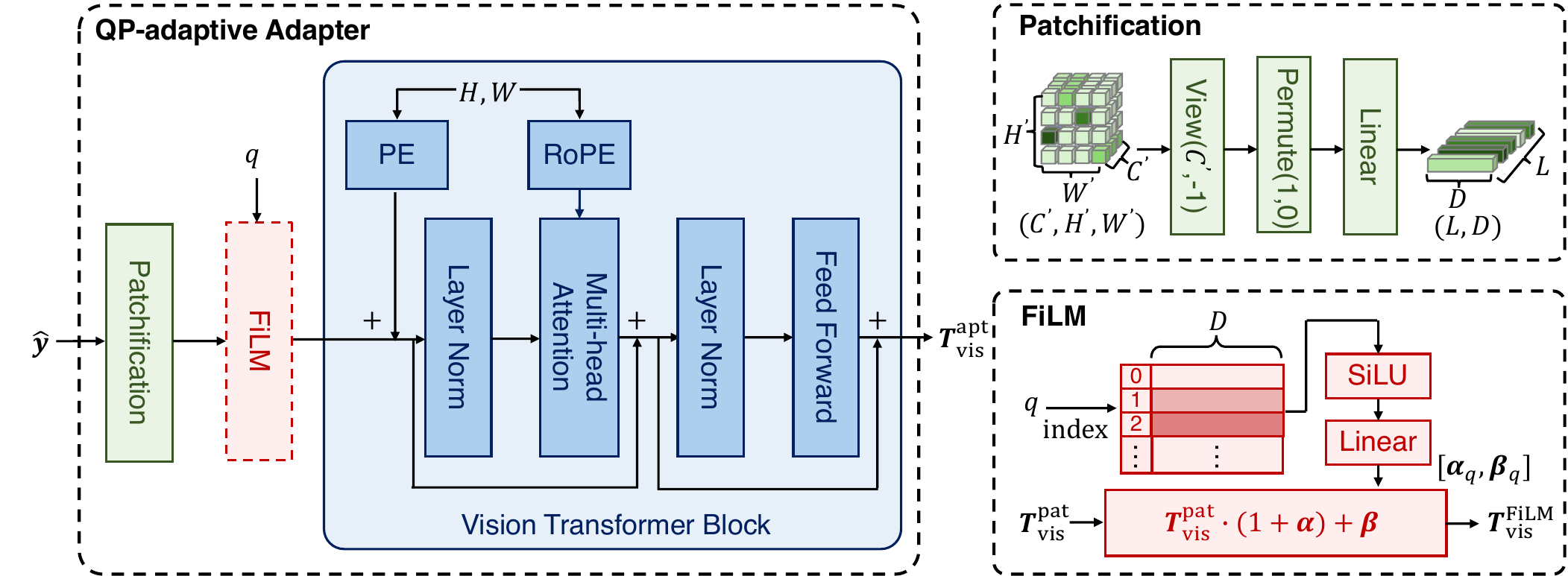}
    \caption{The model architecture of proposed QP-adaptive adapter. It mainly consists of the patchification module, the FiLM module, and the Vision Transformer block. The dashed border of the FiLM block indicates that this module can be skipped. In this case, the adapter converts latents into visual tokens for a fixed QP.}
    \label{fig:adapter}
\end{figure*}

\section{Model Design and Training Scheme}
\label{sec:Model_Design_and_Training_Scheme}

This section presents the architecture design of the proposed adapter first.
The training methods for the neural codec and the adapter are then introduced.
The neural codec is trained independently based on the classical rate-distortion loss.
The adapter is trained based on effective semantic feature extraction for text and images, together with a two-staged semantic alignment loss.

\subsection{Adapter Architecture Design}
In this paper, the receiver needs to convert the compression-oriented latent representation $\hat{\boldsymbol{y}}\in\mathbb{R}^{C'\times H'\times W'}$ into visual token representations $\boldsymbol{T}^{\operatorname{apt}}_{\operatorname{vis}}\in\mathbb{R}^{L\times D}$ suitable for MLLMs.
The proposed adapter performs this conversion through a patchification operation and a Vision Transformer block.
To adapt the adapter to different QPs of the image neural codec, the FiLM module can be further introduced to incorporate the QP information.
The architecture of the proposed adapter is shown in Fig. \ref{fig:adapter}.
The only difference between the QP-adaptive adapter and the QP-fixed adapter is whether the FiLM module is used.
Therefore, the following description focuses on the QP-adaptive adapter.

The patchification operation first converts the two-dimensional (2D) grid feature $\hat{\boldsymbol{y}}$ produced by the convolutional neural network (CNN)-based codec into a one-dimensional (1D) token sequence. Specifically, each spatial grid is treated as one patch and the patchified visual tokens are obtained by
\begin{align}
\boldsymbol{T}^{\operatorname{pat}}_{\operatorname{vis}}
=
\operatorname{Linear}_{\operatorname{pat}}
\left(
\operatorname{Reshape}\left(\hat{\boldsymbol{y}}\right);
\boldsymbol{\theta}_{\operatorname{pat}}
\right),
\end{align}
where $\operatorname{Reshape}(\cdot)$ converts $\hat{\boldsymbol{y}}\in\mathbb{R}^{C'\times H'\times W'}$ into a latent sequence in $\mathbb{R}^{L\times C'}$.
$\operatorname{Linear}_{\operatorname{pat}}(\cdot)$ maps the latent dimension $C'$ to the hidden dimension $D$ of the vision tokenizer.
$\boldsymbol{\theta}_{\operatorname{pat}}$ denotes the learnable parameters of this linear projection.
The output $\boldsymbol{T}^{\operatorname{pat}}_{\operatorname{vis}}\in\mathbb{R}^{L\times D}$ denotes the visual token sequence after patchification, where $L=H'\times W'$ is the number of visual tokens corresponding to the input image.

The FiLM module fuses the QP information with the patchified visual token sequence.
For a given QP index $q$, the QP-dependent modulation parameters are generated as
\begin{align}
\left[
\boldsymbol{\alpha}_{q},
\boldsymbol{\beta}_{q}
\right]
=
\operatorname{Linear}_{\operatorname{qp}}
\left(
\operatorname{SiLU}
\left(
\operatorname{Emb}_{\operatorname{qp}}[q]
\right);
\boldsymbol{\theta}_{\operatorname{qp}}
\right),
\end{align}
where $\operatorname{Emb}_{\operatorname{qp}}[q]\in\mathbb{R}^{1\times D}$ denotes the embedding vector indexed by $q$ from the learnable QP embedding table.
Here, $\operatorname{SiLU}(\cdot)$ denotes the sigmoid linear unit (SiLU) activation.
The operator $\operatorname{Linear}_{\operatorname{qp}}(\cdot)$ maps the QP embedding to a $2D$-dimensional vector, which is split into the QP-dependent scaling and shifting parameters $\boldsymbol{\alpha}_{q},\boldsymbol{\beta}_{q}\in\mathbb{R}^{1\times D}$.

Based on $\boldsymbol{\alpha}_{q},\boldsymbol{\beta}_{q}$, FiLM-modulated visual tokens can be obtained by
\begin{align}
\boldsymbol{T}^{\operatorname{FiLM}}_{\operatorname{vis}}
=
\boldsymbol{T}^{\operatorname{pat}}_{\operatorname{vis}}
\odot
\left(
\boldsymbol{1}_{L}
\left(
\boldsymbol{1}_{D}
+
\boldsymbol{\alpha}_{q}
\right)
\right)
+
\boldsymbol{1}_{L}\boldsymbol{\beta}_{q},
\end{align}
where $\boldsymbol{T}^{\operatorname{FiLM}}_{\operatorname{vis}}\in\mathbb{R}^{L\times D}$ denotes the output of the FiLM module.
Here, $\boldsymbol{1}_{L}\in\mathbb{R}^{L\times 1}$ and $\boldsymbol{1}_{D}\in\mathbb{R}^{1\times D}$ are the all-one vectors.
With the FiLM module, the same adapter can be used for different QPs during inference.
The receiver only needs to provide the current QP index to the adapter.
For a QP-fixed adapter, the FiLM module is skipped, namely $\boldsymbol{T}^{\operatorname{FiLM}}_{\operatorname{vis}}=\boldsymbol{T}^{\operatorname{pat}}_{\operatorname{vis}}$.

To improve the quality of the one-dimensional visual token sequence while controlling the model complexity, we use only one Vision Transformer block in the adapter.
At the beginning of the block, positional encoding (PE) is added to the FiLM-modulated tokens:
\begin{align}
\boldsymbol{T}^{0}_{\operatorname{vis}}
=
\boldsymbol{T}^{\operatorname{FiLM}}_{\operatorname{vis}}
+
\boldsymbol{E}_{\operatorname{PE}}\left(H',W'\right),
\end{align}
where $\boldsymbol{E}_{\operatorname{PE}}\left(H',W'\right)\in\mathbb{R}^{L\times D}$ denotes the interpolated positional embedding.
Rotary position embedding (RoPE) \cite{su2024roformer} is further used in the multi-head attention module to encode 2D spatial positions. 
Different from the learnable PE, RoPE does not introduce additional learnable parameters. 
In the multi-head attention module, it is applied to the query and key vectors before computing the attention scores.

Finally, the Vision Transformer block can be formulated as
\begin{align}
\bar{\boldsymbol{T}}_{\operatorname{vis}}
=
\boldsymbol{T}^{0}_{\operatorname{vis}}
+
\operatorname{MHA}
\left(
\operatorname{LN}
\left(
\boldsymbol{T}^{0}_{\operatorname{vis}}
\right);
\boldsymbol{C}_{\operatorname{RoPE}},
\boldsymbol{S}_{\operatorname{RoPE}},
\boldsymbol{\theta}_{\operatorname{attn}}
\right),
\end{align}
\begin{align}
\boldsymbol{T}^{\operatorname{apt}}_{\operatorname{vis}}
=
\bar{\boldsymbol{T}}_{\operatorname{vis}}
+
\operatorname{FFN}
\left(
\operatorname{LN}
\left(
\bar{\boldsymbol{T}}_{\operatorname{vis}}
\right);
\boldsymbol{\theta}_{\operatorname{ffn}}
\right),
\end{align}
where $\operatorname{LN}(\cdot)$ denotes layer normalization, $\operatorname{MHA}(\cdot)$ denotes multi-head attention, and $\operatorname{FFN}(\cdot)$ denotes the feed-forward network based on multilayer perceptron.
The parameters $\boldsymbol{\theta}_{\operatorname{attn}}$ and $\boldsymbol{\theta}_{\operatorname{ffn}}$ are learnable parameters. $\boldsymbol{C}_{\operatorname{RoPE}}$ and $\boldsymbol{S}_{\operatorname{RoPE}}$ denote the cosine and sine components used in RoPE.

\subsection{Training Scheme for the Neural Codec} 
We now introduce the training scheme in this paper.
For the framework presented in Section \ref{sec:Overview_of_the_Proposed_Token_Communication_framework_for_MLLMs}, the trainable components include the image neural codec and the adapter.
The image neural codec consists of the encoder at the transmitter, the decoder at the receiver, and the entropy model required for entropy coding.
The image neural codec and the adapter are trained separately.
In this subsection, we first introduce the training scheme for the image neural codec.

For the image neural codec, we follow the widely used rate-distortion optimization strategy in neural image compression \cite{balle2018variational, jia2025towards, he2022elic}.
The training objective is defined as
\begin{align}
\mathcal{L}_{\operatorname{codec}}
=
R\left(\hat{\boldsymbol{y}}\right)
+
\lambda_{\operatorname{rd}}
D\left(
\boldsymbol{x},
\hat{\boldsymbol{x}}
\right),
\end{align}
where $R(\hat{\boldsymbol{y}})$ denotes the rate term.
It measures the number of bits required to entropy-code the quantized latent symbols $\hat{\boldsymbol{y}}$.
The distortion between the original image and the reconstructed image is denoted by $D(\boldsymbol{x},\hat{\boldsymbol{x}})$.
The coefficient $\lambda_{\operatorname{rd}}$ controls the tradeoff between rate and distortion.

In hyperprior-based entropy modeling, the rate term is commonly estimated as
\begin{align}
R\left(\hat{\boldsymbol{y}}\right)
=
\mathbb{E}_{\boldsymbol{x}\sim p_{\boldsymbol{x}}}
\left[
-\log_{2}
P_{\hat{\boldsymbol{y}}\mid \hat{\boldsymbol{z}}}
\left(
\hat{\boldsymbol{y}}
\mid
\hat{\boldsymbol{z}}
\right)
-
\log_{2}
P_{\hat{\boldsymbol{z}}}
\left(
\hat{\boldsymbol{z}}
\right)
\right],
\end{align}
where $p_{\boldsymbol{x}}$ denotes the distribution of training images.
The variable $\hat{\boldsymbol{z}}$ denotes the quantized hyperprior.
It is obtained by applying the hyper-analysis transform in the entropy model $\Phi(~\cdot~;\boldsymbol{\theta}_{\operatorname{entropy}})$ to $\boldsymbol{y}$ and then performing quantization.
The conditional probability $p_{\hat{\boldsymbol{y}}\mid \hat{\boldsymbol{z}}}(\hat{\boldsymbol{y}}\mid \hat{\boldsymbol{z}})$ and the hyperprior probability $p_{\hat{\boldsymbol{z}}}(\hat{\boldsymbol{z}})$ are estimated by the entropy model \cite{balle2018variational}.
Therefore, the expected coding length of the quantized latents can be computed during training.

The distortion term $D(\boldsymbol{x},\hat{\boldsymbol{x}})$ can be implemented by the mean squared error (MSE), the absolute error, or other perceptual losses based on reconstruction requirements.
In this paper, we adopt the MSE loss.
The reason is that the reconstructed image at the receiver is used as a reconstruction prior for the MLLM.
Thus, it is expected to preserve texture and structural details from the original image.
Such pixel-level information can better complement the visual tokens produced by the semantically aligned adapter.

\begin{figure*}[!t]
    \centering
    \includegraphics[width=0.8\textwidth]{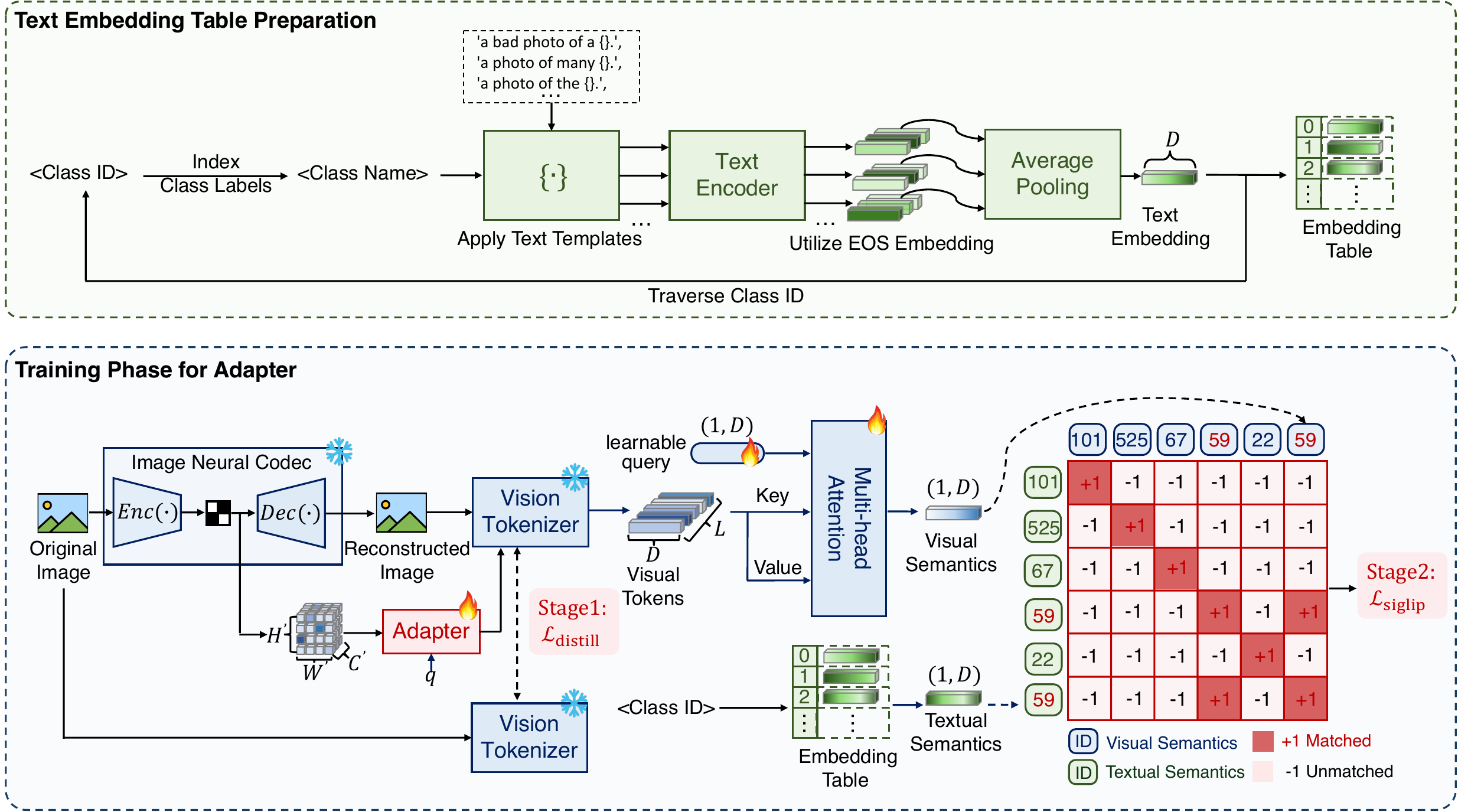}
    \caption{The illustration of proposed training scheme for adapter. Before adapter training, a text embedding table is first prepared. This table allows the corresponding textual semantic feature to be indexed by the class ID of each image during training. The adapter is then trained in two stages. In the first stage, a distillation-based loss function is adopted. In the second stage, a SigLIP-based loss is used for semantic alignment.}
    \label{fig:train}
\end{figure*}

\subsection{Training Scheme for the Adapter}
This subsection introduces the training scheme for the adapter.
As shown in Fig. \ref{fig:train}, the proposed scheme consists of a text embedding table preparation process and a two-stage adapter training process.
Before training the adapter, a text embedding table is constructed to provide textual semantic features for different image classes.
During adapter training, the first stage uses a distillation loss to align the intermediate output of the vision tokenizer.
The second stage uses a semantic alignment loss based on SigLIP to align the visual semantics with the corresponding textual semantics.
In the whole training process, the image neural codec, the vision tokenizer, and the text encoder are frozen, and only the adapter and the attention pooling head are optimized.

For text embedding table preparation, we denote $\mathcal{C}=\{0,1,\cdots,N_{\operatorname{cls}}-1\}$ as the set of class IDs in the training dataset.
For the $c$-th class, its class name is inserted into $M$ text templates to generate a set of textual descriptions for this class of images as
\begin{align}
\mathcal{S}_{c}
=
\left\{
s_{c}^{m}
=
\operatorname{Temp}_{m}
\left(
\operatorname{Label}(c)
\right)
\right\}_{m=1}^{M},
\end{align}
where $\operatorname{Label}(c)$ denotes the class name of class $c$, and $\operatorname{Temp}_{m}(\cdot)$ denotes the $m$-th text template.
These textual descriptions are fed into the text encoder paired with the continuous vision tokenizer used by the MLLM.
For each description $s_{c}^{m}$, the text encoder produces a sequence of text token embeddings:
\begin{align}
\boldsymbol{G}_{c}^{m}
=
\operatorname{TextEnc}
\left(
s_{c}^{m};
\boldsymbol{\theta}_{\operatorname{text}}
\right)
\in
\mathbb{R}^{L_{\operatorname{text}}\times D},
\end{align}
where $\boldsymbol{\theta}_{\operatorname{text}}$ denotes the parameters of the frozen text encoder, and $L_{\operatorname{text}}$ denotes the text token sequence length.
We use the embedding corresponding to the end-of-sequence (EOS) token as the semantic embedding of this description:
\begin{align}
\boldsymbol{g}_{c}^{m}
=
\operatorname{Norm}
\left(
\boldsymbol{G}_{c}^{m}
\left[
\operatorname{EOS}
\right]
\right)
\in
\mathbb{R}^{1\times D},
\end{align}
where $\operatorname{Norm}(\cdot)$ denotes $\ell_{2}$ normalization.
The text embedding of class $c$ is obtained by average pooling over all template-based descriptions:
\begin{align}
\boldsymbol{t}_{c}
=
\operatorname{Norm}
\left(
\frac{1}{M}
\sum_{m=1}^{M}
\boldsymbol{g}_{c}^{m}
\right)
\in
\mathbb{R}^{1\times D}.
\end{align}
By traversing all classes in the training dataset, the text embedding table is constructed as
\begin{align}
\boldsymbol{E}_{\operatorname{text}}
=
\left[
\boldsymbol{t}_{0}^{T},
\boldsymbol{t}_{1}^{T},
\cdots,
\boldsymbol{t}_{N_{\operatorname{cls}}-1}^{T}
\right]^{T}
\in
\mathbb{R}^{N_{\operatorname{cls}}\times D}.
\end{align}
During adapter training, the textual semantic feature corresponding to an image with class ID $c_i$ can be directly obtained by indexing $\boldsymbol{E}_{\operatorname{text}}[c_i]$.

For adapter training, in the first training stage, we use a distillation loss to warm up the adapter.
For an input image $\boldsymbol{x}_{i}$, the frozen neural codec produces the quantized latent representation $\hat{\boldsymbol{y}}_{i}$ and the reconstructed image $\hat{\boldsymbol{x}}_{i}$.
The reconstructed image is fed into the vision tokenizer and processed by the first $n-1$ Transformer blocks to obtain
\begin{align}
\boldsymbol{T}_{\operatorname{vis},i}^{\operatorname{rec},n-1}
=
\mathcal{V}_{1:n-1}
\left(
\operatorname{Tok}_{\operatorname{vis}}
\left(
\hat{\boldsymbol{x}}_{i}
\right);
\boldsymbol{\theta}_{\mathcal{V}_{1:n-1}}
\right),
\end{align}
where $\operatorname{Tok}_{\operatorname{vis}}(\cdot)$ denotes the patchification and embedding projection of the vision tokenizer, and $\mathcal{V}_{1:n-1}(\cdot)$ denotes the first $n-1$ Transformer blocks.
Meanwhile, the adapter converts $\hat{\boldsymbol{y}}_{i}$ into visual tokens $\boldsymbol{T}_{\operatorname{vis},i}^{\operatorname{apt}}$.
The two token sequences are fused by addition:
\begin{align}
\boldsymbol{T}_{\operatorname{vis},i}^{\operatorname{fuse},n-1}
=
\boldsymbol{T}_{\operatorname{vis},i}^{\operatorname{rec},n-1}
+
\alpha
\boldsymbol{T}_{\operatorname{vis},i}^{\operatorname{apt}},
\label{fusion}
\end{align}
where $\alpha$ controls the injection strength of the adapter output.
The original image is also processed by the same frozen vision tokenizer to provide the distillation target:
\begin{align}
\boldsymbol{T}_{\operatorname{vis},i}^{\operatorname{ori},n-1}
=
\mathcal{V}_{1:n-1}
\left(
\operatorname{Tok}_{\operatorname{vis}}
\left(
\boldsymbol{x}_{i}
\right);
\boldsymbol{\theta}_{\mathcal{V}_{1:n-1}}
\right).
\end{align}
The stage-one distillation loss is defined as
\begin{align}
\mathcal{L}_{\operatorname{distill}}
=
\frac{1}{|\mathcal{B}|LD}
\sum_{i\in\mathcal{B}}
\left\|
\boldsymbol{T}_{\operatorname{vis},i}^{\operatorname{fuse},n-1}
-
\operatorname{sg}
\left(
\boldsymbol{T}_{\operatorname{vis},i}^{\operatorname{ori},n-1}
\right)
\right\|^{2},
\label{distill}
\end{align}
where $\mathcal{B}$ denotes a mini-batch, $|\mathcal{B}|$ denotes the batch size, and $\operatorname{sg}(\cdot)$ denotes the stop-gradient operation.
This loss encourages the injected visual tokens to compensate for the information loss caused by compression at the intermediate layer of the vision tokenizer.

In the second training stage, semantic alignment is performed between the visual semantics and the textual semantics.
The fused intermediate tokens are further processed by the remaining Transformer blocks of the frozen vision tokenizer:
\begin{align}
\boldsymbol{T}_{\operatorname{vis},i}^{\operatorname{out}}
=
\mathcal{V}_{n:N_{\mathcal{V}}}
\left(
\boldsymbol{T}_{\operatorname{vis},i}^{\operatorname{fuse},n-1};
\boldsymbol{\theta}_{\mathcal{V}_{n:N_{\mathcal{V}}}}
\right),
\end{align}
where $\mathcal{V}_{n:N_{\mathcal{V}}}(\cdot)$ denotes the Transformer blocks from the $n$-th block to the final block of the vision tokenizer.
A learnable query is then used to aggregate the visual token sequence into a visual semantic feature through attention pooling:
\begin{align}
\boldsymbol{v}_{i}
=
\operatorname{Norm}
\left(
\operatorname{MHA}_{\operatorname{pool}}
\left(
\boldsymbol{q}_{\operatorname{pool}},
\boldsymbol{T}_{\operatorname{vis},i}^{\operatorname{out}},
\boldsymbol{T}_{\operatorname{vis},i}^{\operatorname{out}};
\boldsymbol{\theta}_{\operatorname{pool}}
\right)
\right),
\label{visual_semantic}
\end{align}
where $\boldsymbol{q}_{\operatorname{pool}}\in\mathbb{R}^{1\times D}$ is a learnable query, and $\operatorname{MHA}_{\operatorname{pool}}(\cdot)$ denotes the multi-head attention pooling module. 
Here, $\boldsymbol{q}_{\operatorname{pool}}$ is the query of the attention pooling module, while $\boldsymbol{T}_{\operatorname{vis},i}^{\operatorname{out}}$ is used as both the key and value. 
The textual semantic feature of image $\boldsymbol{x}_{i}$ with class ID $c_i$ is indexed from the prepared text embedding table:
\begin{align}
\boldsymbol{u}_{i}
=
\boldsymbol{E}_{\operatorname{text}}[c_i].
\label{textual_semantic}
\end{align}

\begin{algorithm}[!t]
\caption{Training the Proposed Adapter}
\label{alg:adapter_training}
\begin{algorithmic}[1]
\State \textbf{Input:}
Training dataset $\mathcal{D}=\{(\boldsymbol{x}_{i},c_i)\}$, trained neural codec $\boldsymbol{\theta}^{*}_{\operatorname{enc}}$, $\boldsymbol{\theta}^{*}_{\operatorname{dec}}$, $\boldsymbol{\theta}^{*}_{\operatorname{entropy}}$, text embedding table $\boldsymbol{E}_{\operatorname{text}}$, QP index $q$, training steps $N_{\operatorname{dis}}$ and $N_{\operatorname{align}}$.
\State \textbf{Output:}
Trained adapter parameters $\boldsymbol{\theta}^{*}_{\operatorname{apt}}(q)$.

\State \textbf{Initialization:} Initialize adapter $\boldsymbol{\theta}^{*}_{\operatorname{apt}}(q)$, pooling head $\boldsymbol{\theta}^{*}_{\operatorname{pool}}(q)$, and SigLIP parameters $\tau^{*}$ and $b^{*}$.

\State \textbf{Stage 1: Distillation Warm-up}
\For{$i \leftarrow 1$ to $N_{\operatorname{dis}}$}
    \State Sample a mini-batch $\mathcal{B}$ from $\mathcal{D}$.
    \State Obtain $\hat{\boldsymbol{y}}$, $\hat{\boldsymbol{x}}$, $\boldsymbol{T}_{\operatorname{vis}}^{\operatorname{apt}}$, $\boldsymbol{T}_{\operatorname{vis}}^{\operatorname{rec},n-1}$, and $\boldsymbol{T}_{\operatorname{vis}}^{\operatorname{ori},n-1}$ under $q$.
    \State Compute $\boldsymbol{T}_{\operatorname{vis}}^{\operatorname{fuse},n-1}$ by \eqref{fusion}.
    \State Compute $\mathcal{L}_{\operatorname{distill}}$ by \eqref{distill}.
    \State Update $\boldsymbol{\theta}_{\operatorname{apt}}(q)$ based on $\mathcal{L}_{\operatorname{distill}}$.
\EndFor

\State \textbf{Stage 2: Semantic Alignment}
\For{$i \leftarrow 1$ to $N_{\operatorname{align}}$}
    \State Compute $\mathcal{L}_{\operatorname{distill}}$ following Stage 1.
    \State Extract visual semantic features $\boldsymbol{v}_{i}$ by \eqref{visual_semantic}.
    \State Index textual semantic features by \eqref{textual_semantic}.
    \State Compute $\mathcal{L}_{\operatorname{SigLIP}}$ by \eqref{siglip}.
    \State Update $\boldsymbol{\theta}_{\operatorname{apt}}(q)$, $\boldsymbol{\theta}_{\operatorname{pool}}(q)$, $\tau$, $b$ based on $\mathcal{L}_{\operatorname{distill}}+\lambda_{\operatorname{align}}\mathcal{L}_{\operatorname{SigLIP}}$.
\EndFor
\end{algorithmic}
\end{algorithm}

Based on the above formulation, we build the proposed semantic alignment loss.
For a pair of visual and textual semantic features, the goal is to increase the similarity of matched pairs and decrease the similarity of unmatched pairs.
Let $\ell_{ij}$ denote the similarity logit between the visual semantic feature $\boldsymbol{v}_{i}$ and the textual semantic feature $\boldsymbol{u}_{j}$.
For a binary matching label $m_{ij}\in\{0,1\}$, the standard binary cross-entropy loss can be written as
\begin{align}
\mathcal{L}_{\operatorname{BCE}}^{ij}
=
-
m_{ij}\log\sigma\left(\ell_{ij}\right)
-
\left(1-m_{ij}\right)
\log
\left(
1-\sigma\left(\ell_{ij}\right)
\right),
\end{align}
where $\sigma(\cdot)$ denotes the sigmoid function.
To express matched and unmatched pairs in a unified form, we convert the binary label $m_{ij}\in\{0,1\}$ into a signed label $z_{ij}\in\{+1,-1\}$:
\begin{align}
z_{ij}
=
2m_{ij}-1.
\end{align}
Here, $z_{ij}=+1$ denotes a matched image-text pair, and $z_{ij}=-1$ denotes an unmatched pair.
With this signed label, the binary loss can be equivalently written as
\begin{align}
\mathcal{L}_{\operatorname{binary}}^{ij}
=
-\log
\sigma
\left(
z_{ij}\ell_{ij}
\right).
\end{align}
This form gives $-\log\sigma(\ell_{ij})$ for matched pairs and $-\log\sigma(-\ell_{ij})$ for unmatched pairs. The equivalence for matched pairs is straightforward. For unmatched pairs, it can be proved that  $\sigma(-\ell_{ij})=1-\sigma(\ell_{ij})$.

To step further, we calculate the similarity logit as
\begin{align}
\ell_{ij}
=
\tau
\boldsymbol{v}_{i}
\boldsymbol{u}_{j}^{T}
+
b,
\end{align}
where $\tau$ and $b$ are learnable scale and bias, respectively. Different from the original pairwise setting that only treats diagonal image-text pairs as positives \cite{zhai2023sigmoid}, images and text features with the same class ID are treated as positives in our implementation, which means
\begin{align}
z_{ij}
=
2\cdot\mathbbm{1}\left(c_i=c_j\right)-1.
\end{align}

Therefore, the proposed SigLIP loss over the mini-batch is formulated as
\begin{align}
\mathcal{L}_{\operatorname{SigLIP}}
=
-\frac{1}{|\mathcal{B}|}
\sum_{i=1}^{|\mathcal{B}|}
\sum_{j=1}^{|\mathcal{B}|}
\log
\sigma
\left(
z_{ij}
\left(
\tau
\boldsymbol{v}_{i}
\boldsymbol{u}_{j}^{T}
+
b
\right)
\right).
\label{siglip}
\end{align}
Although the second stage primarily targets semantic alignment, its total objective still retains the distillation term to prevent undesired drift in the adapter parameters:
\begin{align}
\mathcal{L}_{\operatorname{stage2}}
=
\mathcal{L}_{\operatorname{distill}}
+
\lambda_{\operatorname{align}}
\mathcal{L}_{\operatorname{SigLIP}},
\end{align}
where $\lambda_{\operatorname{align}}$ controls the strength of semantic alignment.

For a fixed quantization QP, the above two-stage training scheme gives a QP-fixed adapter. That is, each QP of the neural codec corresponds to one fixed-QP adapter. The training process is concluded in Algorithm \ref{alg:adapter_training}.
For the QP-adaptive adapter, training from scratch is unnecessary.
We initialize it with the fixed-QP adapter trained at a middle-rate point within the available QP range.
During QP-adaptive training, one QP is randomly sampled for each image in the mini-batch.
The sampled QP $q_i$ is used by both the neural codec and the FiLM-based adapter.
It is worthy to note that the original image branch and the attention pooling head are only used during training.
They are not required during inference.
During inference, the receiver only needs the reconstructed image, the decoded latent representation, the current QP, and the trained adapter to produce visual tokens for the MLLM.

\section{Simulation Results}
\label{sec:Simulation_Results}

This section presents the detailed simulation settings, including the adopted image neural codec, MLLMs, training dataset, and evaluation benchmarks.
In addition, detailed performance curves, visualization results, and the corresponding analyses are provided.

\subsection{Simulation Settings}

\subsubsection{Training Settings} For the image neural codec, we construct the training dataset by combining high-resolution images from ImageNet whose height and width are both larger than 500 pixels with the DIV2K and Flickr2K datasets.
During training, each image is randomly cropped into a $256\times256$ patch.
Two neural codecs, namely DCVC-RT \cite{jia2025towards} and ELIC \cite{he2022elic} are trained.
The batch size is set to $16$.
The training process is divided into two stages.
In the first stage, the straight-through estimator (STE) is used to approximate the quantization process, and the learning rate is set to $1\times10^{-5}$.
After the loss converges, the second stage uses additive uniform noise to approximate the quantization process with learning rate $1\times10^{-6}$.
Other training details follow the original settings \cite{jia2025towards}, \cite{he2022elic}.
It is worth noting that the official checkpoint of DCVC-RT cannot be directly used because it is trained with images in the YUV format, which is incompatible with the vision tokenizers commonly used by MLLMs.

For the adapter, we use the complete ImageNet training dataset.
During training, we resize each image while preserving its original aspect ratio so that the shorter side becomes $256$ pixels.
Then, the center $256\times256$ region of the resized image is cropped.
To reduce the training time of each epoch, the whole training set is randomly divided into $12$ subsets, and these subsets are used cyclically during training.
The batch size is set to $256$, and the learning rate is set to $1\times10^{-4}$.
The Adam optimizer is used with the weight decay set to $0$.
$\lambda_{\operatorname{align}}$ follows a ramp-up schedule.
Specifically, it is set to $0$ before the third epoch, linearly increased from $0$ to $10^{-3}$ between the third and tenth epochs, and fixed at $10^{-3}$ afterward.
The initial values of the learnable SigLIP parameters $\tau$ and $b$ are set to $10$ and $-10$, respectively, following \cite{zhai2023sigmoid}.
The visual tokens produced by the adapter are injected into the input of the third block of the vision tokenizer.
The injection strength $\alpha$ is set to $0.1$.
As for the construction of text embedding table in adapter training, the text encoder is selected according to the MLLM used in the pipeline.
When Qwen3-VL \cite{bai2025qwen3} is used as the MLLM, the text encoder corresponding to SigLIP2 \cite{tschannen2025siglip} is used to extract textual semantic features. If LLaMA-Adapter \cite{zhang2024llama} is used as the MLLM, the text encoder of Contrastive Language-Image Pretraining (CLIP) \cite{radford2021learning} is selected to extract textual semantic features.

\subsubsection{Testing Settings}
We evaluate the proposed scheme and the baseline methods on four MLLM benchmarks, including MME \cite{fu2025mme}, POPE \cite{li2023evaluating}, SeedBench \cite{li2024seed}, and COCO Caption \cite{karpathy2015deep}, to comprehensively assess the performance of MLLMs under different image processing schemes. For POPE, we use the COCO-POPE popular setting. For COCO Caption, we use the Karpathy test split for evaluation.

For clarity in the following performance analysis, we briefly introduce different image processing schemes for MLLMs and define their notations as follows:
\begin{itemize}
  \item \textit{Uncompressed}: The original images are directly fed into the MLLM.
  This scheme usually provides the upper bound of the performance achieved by different image processing schemes on various benchmarks.

  \item \textit{A-ImageTC}: The proposed Adapter-based Image Token Communication (A-ImageTC) scheme.
  It effectively injects visual tokens into the reconstruction prior through the adapter.
  Unless otherwise specified, the codec used by A-ImageTC is DCVC-RT.
  The variants of A-ImageTC are specified when they are discussed.

  \item \textit{Recon}: The reconstruction-based scheme, which is also the common paradigm used in current MLLM interactions.
  In this scheme, images are compressed at the transmitter and reconstructed at the receiver before being fed into the MLLM.

  \item \textit{Bridge}: An MLLM-oriented coding scheme that uses cross-entropy (CE) loss for semantic alignment and contains only the adapter branch at the receiver \cite{kao2025bridging}.
  It should be noted that, when Qwen3-VL is used, our reproduction shows that its original training method cannot converge.
  Therefore, in the experiments related to Qwen3-VL, we keep its loss function, but adopt the proposed architecture with a reconstruction prior and use the attention pooling head to extract visual features.
\end{itemize}

\begin{figure*}[!t]
    \centering
    % a
    \begin{subfigure}{0.85\textwidth}
        \centering
        \includegraphics[width=\textwidth]{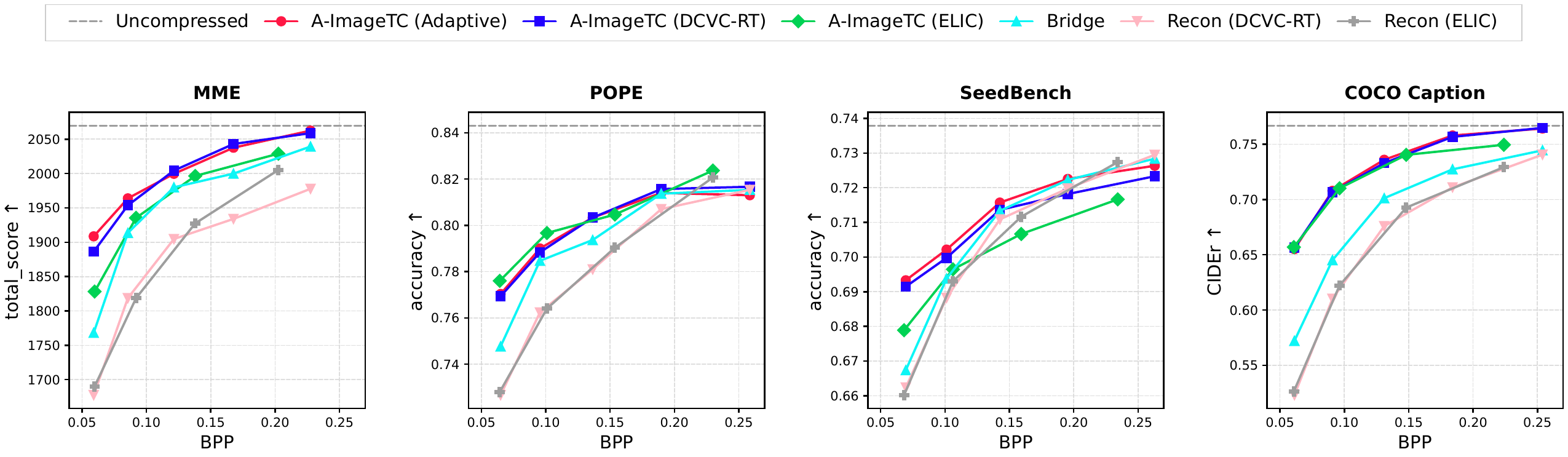}
        \caption{Qwen3-VL-8B}
        \label{fig:performance_8B}
    \end{subfigure}

    % b
    \begin{subfigure}{0.85\textwidth}
        \centering
        \includegraphics[width=\textwidth]{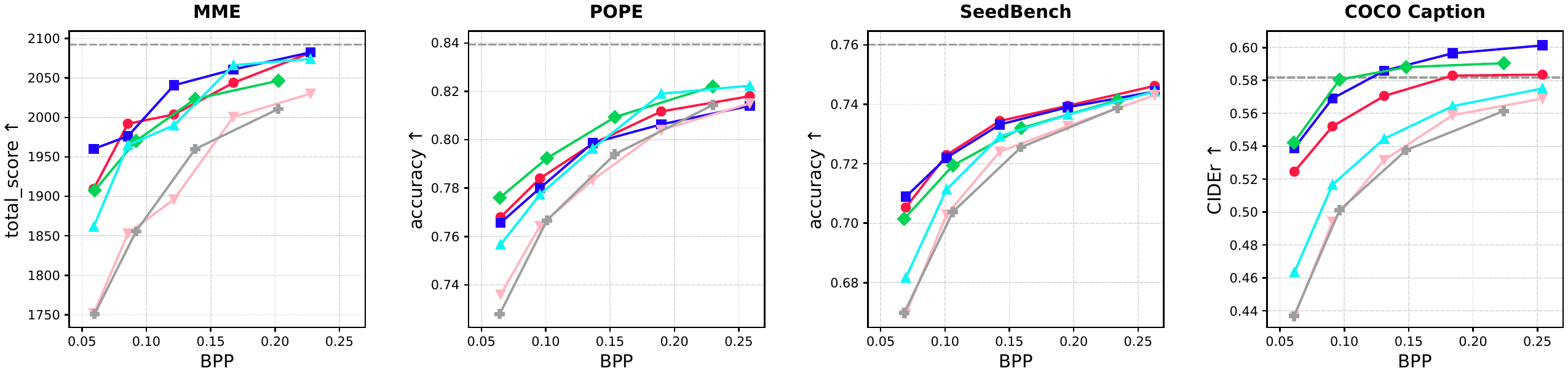}
        \caption{Qwen3-VL-32B}
        \label{fig:performance_32B}
    \end{subfigure}
    
    \caption{Performance of different image processing schemes versus BPP on various MLLM benchmarks. (a) The results with Qwen3-VL-8B. (b) The results with Qwen3-VL-32B.}
    \label{fig:performance}
\end{figure*}

\subsection{Performance Analysis}

\subsubsection{Performance of Proposed Schemes}
Fig. \ref{fig:performance} shows the MLLM performance of different image processing schemes under different BPP.
For Qwen3-VL-8B, the proposed A-ImageTC with DCVC-RT and its QP-adaptive variant achieve better performance than the reconstruction-based schemes and Bridge on most benchmarks and rate points.
The advantage is especially clear in the low-BPP region.
This demonstrates that directly optimizing the reconstructed image for human perception is not always sufficient for MLLM-oriented transmission.
By injecting adapter-generated visual tokens into the vision tokenizer, A-ImageTC can provide more informative visual representations for MLLM inference.

Fig. \ref{fig:performance_8B} also verifies the compatibility of the proposed framework with different neural codecs.
When ELIC is used as the image neural codec, A-ImageTC still outperforms the corresponding reconstruction-based baselines.
For SeedBench, A-ImageTC with ELIC shows a slight performance drop in the high-BPP region.
This may be caused by the limited capacity of ELIC latent representation.
Nevertheless, it still maintains a clear advantage in the low-BPP region.

The results with Qwen3-VL-32B in Fig. \ref{fig:performance_32B} further validate the generality of the proposed scheme.
With the same A-ImageTC framework, increasing the scale of the MLLM generally improves the performance on MME, POPE, and SeedBench.
This indicates that the visual tokens produced by the proposed framework can be effectively used by stronger MLLMs.
An exception is observed on COCO Caption, where Qwen3-VL-32B obtains lower CIDEr scores than Qwen3-VL-8B.
This phenomenon is counter-intuitive.
Moreover, A-ImageTC even surpasses the uncompressed image input on COCO Caption with Qwen3-VL-32B.

\begin{figure}[!t]
    \centering
    \includegraphics[width=0.8\columnwidth]{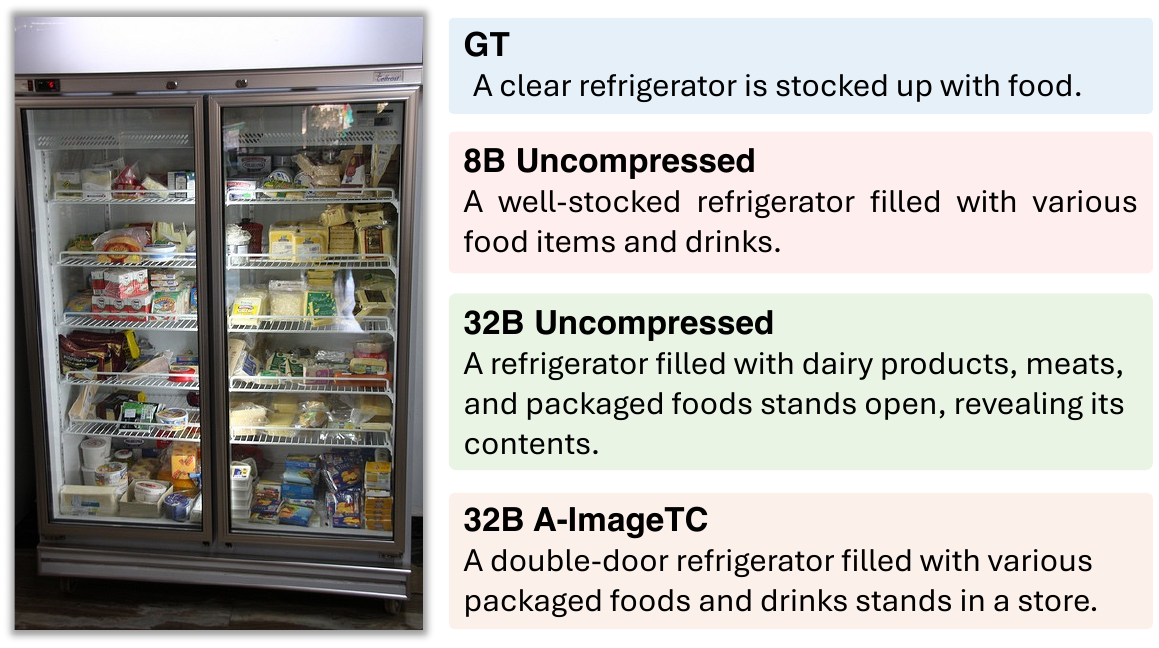}
    \caption{Qualitative example explaining the COCO Caption results for the Qwen3-VL 32B and Qwen3-VL-8B models.}
    \label{fig:performance_32b8b}
\end{figure}

To better understand this observation, we further visualize representative captioning outputs in Fig. \ref{fig:performance_32b8b}.
The caption generated by Qwen3-VL-32B with the uncompressed image contains more detailed descriptions than that generated by Qwen3-VL-8B.
However, the ground-truth caption in COCO Caption is usually concise and focuses only on the most salient objects.
Since CIDEr measures the similarity between generated captions and reference captions based on weighted n-gram matching, a more detailed caption does not necessarily obtain a higher score.
In the shown example, the output of Qwen3-VL-32B with A-ImageTC is closer to the ground-truth caption than the output with the uncompressed image.
Therefore, the higher CIDEr score of A-ImageTC does not imply that compression improves the intrinsic visual understanding ability of the MLLM.

\begin{figure}[!t]
    \centering
    \includegraphics[width=0.85\linewidth]{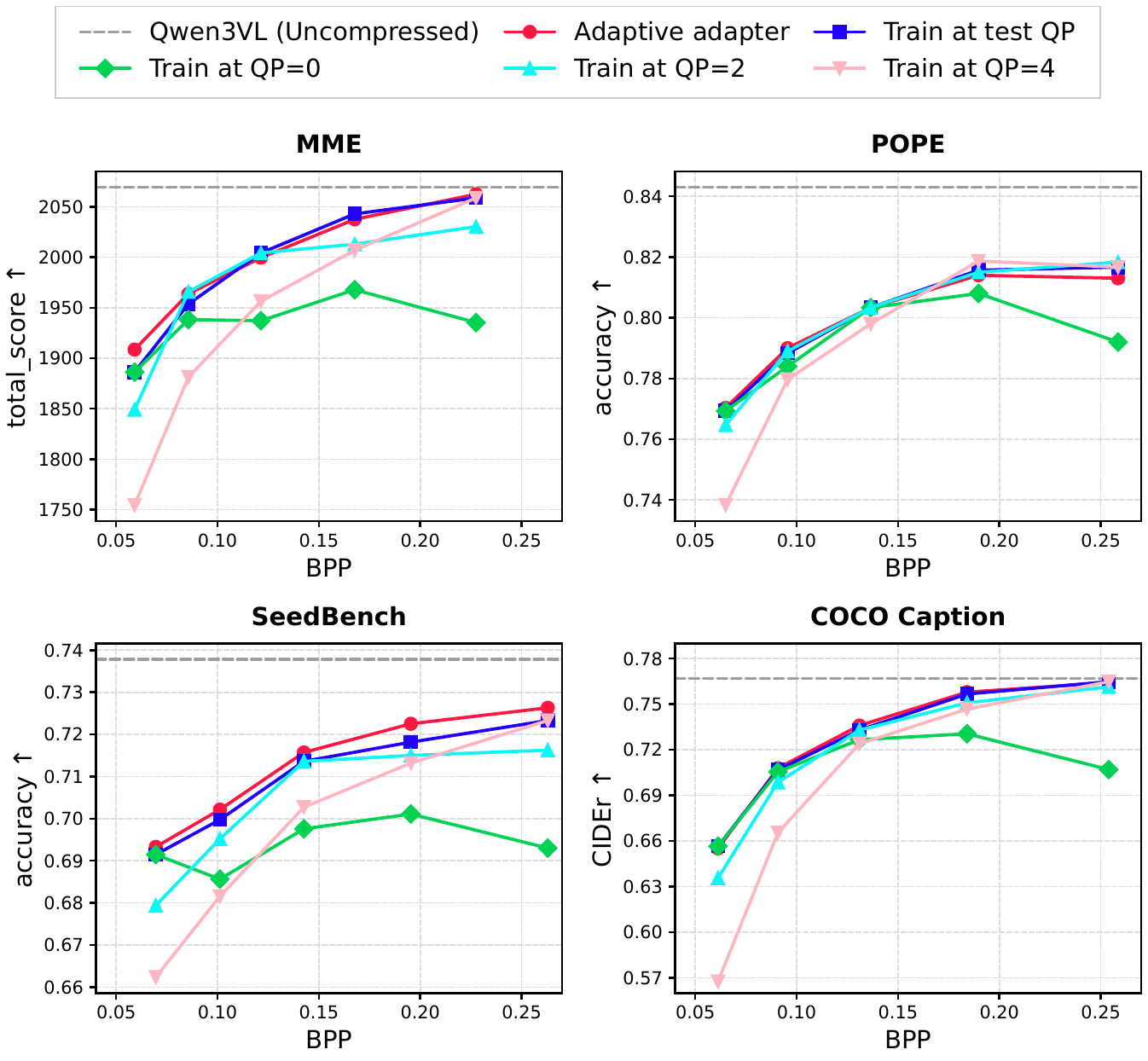}
    \caption{Performance comparison of QP-adaptive adapter and QP-fixed adapters on various MLLM benchmarks.}
    \label{fig:performance_adaptive}
\end{figure}

\begin{figure}[!t]
    \centering
    \includegraphics[width=0.85\linewidth]{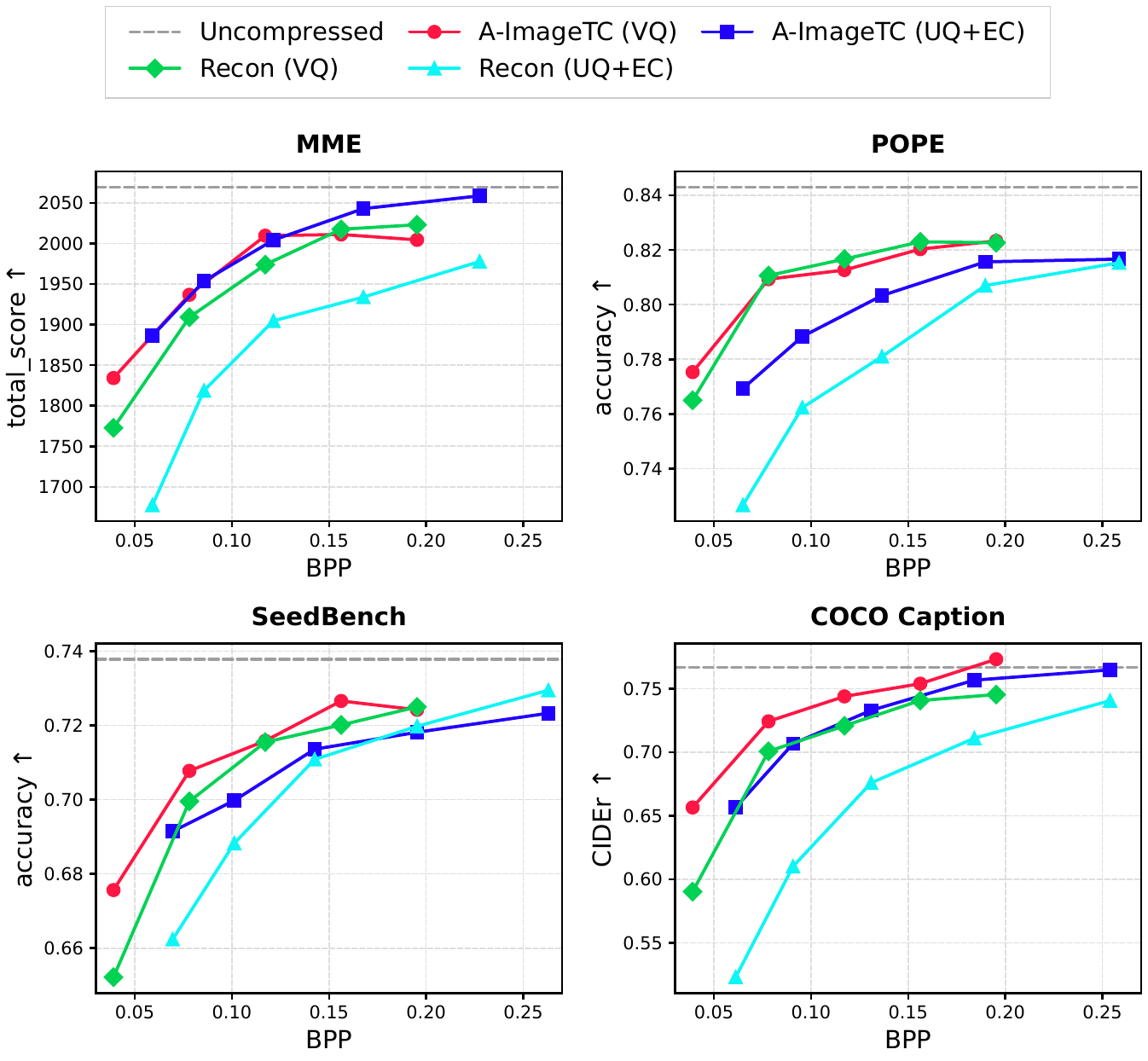}
    \caption{Performance comparison between the proposed scheme and VQ-based token communication methods.}
    \label{fig:performance_UQVQ}
\end{figure}

Fig. \ref{fig:performance_adaptive} evaluates the performance of the proposed QP-adaptive adapter.
It can be observed that, after adaptive training with the FiLM module, a single adapter can effectively convert compression-oriented latents under different QPs into suitable visual tokens.
The QP-adaptive adapter achieves performance comparable to QP-fixed adapters trained at the corresponding test QPs.
It even shows a slight advantage on SeedBench, indicating that training across multiple QPs can improve the robustness of visual token adaptation. 
Fig. \ref{fig:performance_adaptive} also compares adapters trained at a fixed QP and tested under different QPs.
When the test QP deviates from the training QP, clear performance degradation can be observed.
This degradation is especially severe when the adapter trained at QP$=0$ or QP$=4$ is applied to other rates.
This result indicates that the distribution of compression-oriented latents changes with QP, and a fixed-QP adapter cannot consistently handle such distribution shifts.
By incorporating QP information through FiLM, the proposed QP-adaptive adapter alleviates this mismatch and enables one adapter to support multiple codec rates.

Fig. \ref{fig:performance_UQVQ} compares the proposed scheme with VQ-based coding methods.
VQ has been widely adopted in token communication because it produces discrete tokens whose indices naturally provide compact bit representations.
It can generate images with strong perceptual quality at very low bitrates.
In this experiment, we use a VQ-based scheme optimized for perceptual quality \cite{cao2026progic}.
It can be observed that Recon (VQ) performs better than Recon (UQ+EC) and A-ImageTC (UQ+EC) on POPE and SeedBench.
Here, the suffix UQ denotes uniform quantization, and EC denotes entropy coding.
This result indicates that perceptual-oriented optimization can benefit MLLM-oriented tasks when the reconstruction preserves task-relevant visual semantics.
However, VQ can also be integrated into the proposed A-ImageTC framework as a neural codec.
It is shown that A-ImageTC (VQ) further improves the performance of Recon (VQ).
This verifies that the proposed adapter-based token injection is complementary to current VQ-based token communication schemes.

\begin{figure}[!t]
    \centering
    \includegraphics[width=0.65\columnwidth]{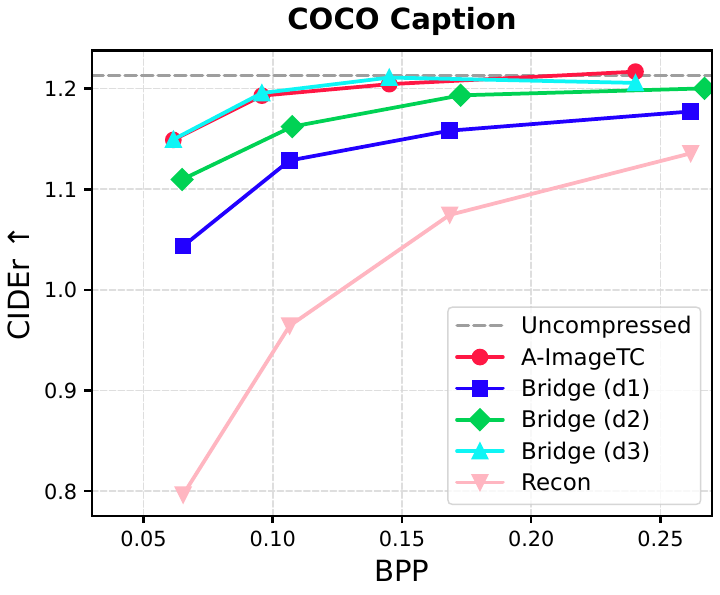}
    \caption{Performance on the COCO Caption task when LLaMA-Adapter is used as the MLLM in the pipeline.}
    \label{fig:performance_bridge}
\end{figure}

\subsubsection{Validation on another MLLM}
To further validate the generality of the proposed method, we conduct experiments with LLaMA-Adapter as the MLLM.
LLaMA-Adapter differs from Qwen3-VL in both the vision tokenizer and the LLM backbone.
Since LLaMA-Adapter is finetuned on specific MLLM visual understanding tasks, we only evaluate the COCO Caption performance in this experiment.

As shown in Fig. \ref{fig:performance_bridge}, the proposed A-ImageTC still achieves a clear advantage over the reconstruction-based scheme and achieves performance comparable to  Bridge (d3) across different BPP values.
This result indicates that the proposed adapter-based token injection is not limited to Qwen3-VL.

\begin{figure}[!t]
    \centering

    \begin{subfigure}{0.48\columnwidth}
        \centering
        \includegraphics[width=\linewidth]{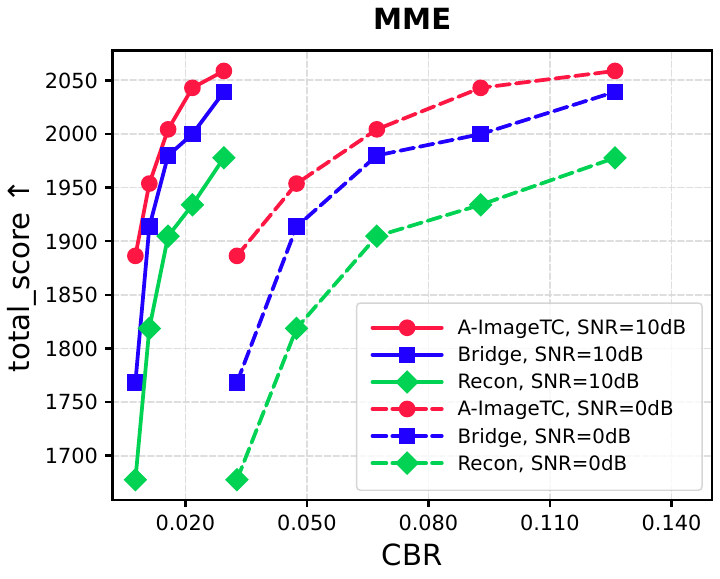}
        \caption{CBR versus MME score}
        \label{fig:cbr}
    \end{subfigure}
    \hfill
    \begin{subfigure}{0.48\columnwidth}
        \centering
        \includegraphics[width=\linewidth]{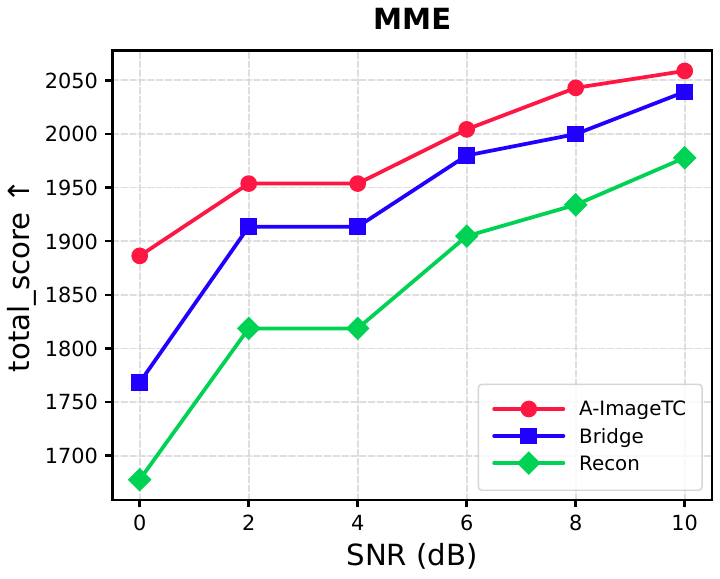}
        \caption{SNR versus MME score}
        \label{fig:snr}
    \end{subfigure}
    
    \caption{MME score of different image processing schemes over Rayleigh fading channels.}
    \label{fig:wireless}
\end{figure}

\subsubsection{Simulation for Wireless Transmission}
We further evaluate the proposed scheme under wireless transmission.
A Rayleigh fading channel is considered, and perfect channel state information (CSI) is assumed to be available for the equalizer.
Since the image is compressed into a bitstream by the neural codec, a classical separated digital transmission pipeline can be adopted.
Specifically, low-density parity-check (LDPC) coding and quadrature amplitude modulation (QAM) are used for channel coding and modulation, respectively.
For different SNRs, MCS is selected based on 3GPP standards \footnote{[Online]. Available: \url{https://www.etsi.org/deliver/etsi_ts/138200_138299/138214/16.02.00_60/ts_138214v160200p.pdf}}.

In Fig. \ref{fig:cbr}, A-ImageTC consistently outperforms Bridge and Recon under the same CBR and SNR.
This result shows that the adapter-generated visual tokens remain beneficial after practical channel coding and modulation. In Fig. \ref{fig:snr},  we set CBR to $0.03$ and compare different schemes under varying SNRs.
As the SNR decreases, the performance of all schemes degrades.
However, A-ImageTC maintains a clear advantage over the baselines across the whole SNR range. The performance loss becomes severe in the low-SNR regime. This indicates that the separated digital transmission pipeline still suffers from the cliff effect under poor channel conditions.

\subsubsection{Visualized Examples}
\begin{figure*}[!t]
    \centering
    \includegraphics[width=0.8\textwidth]{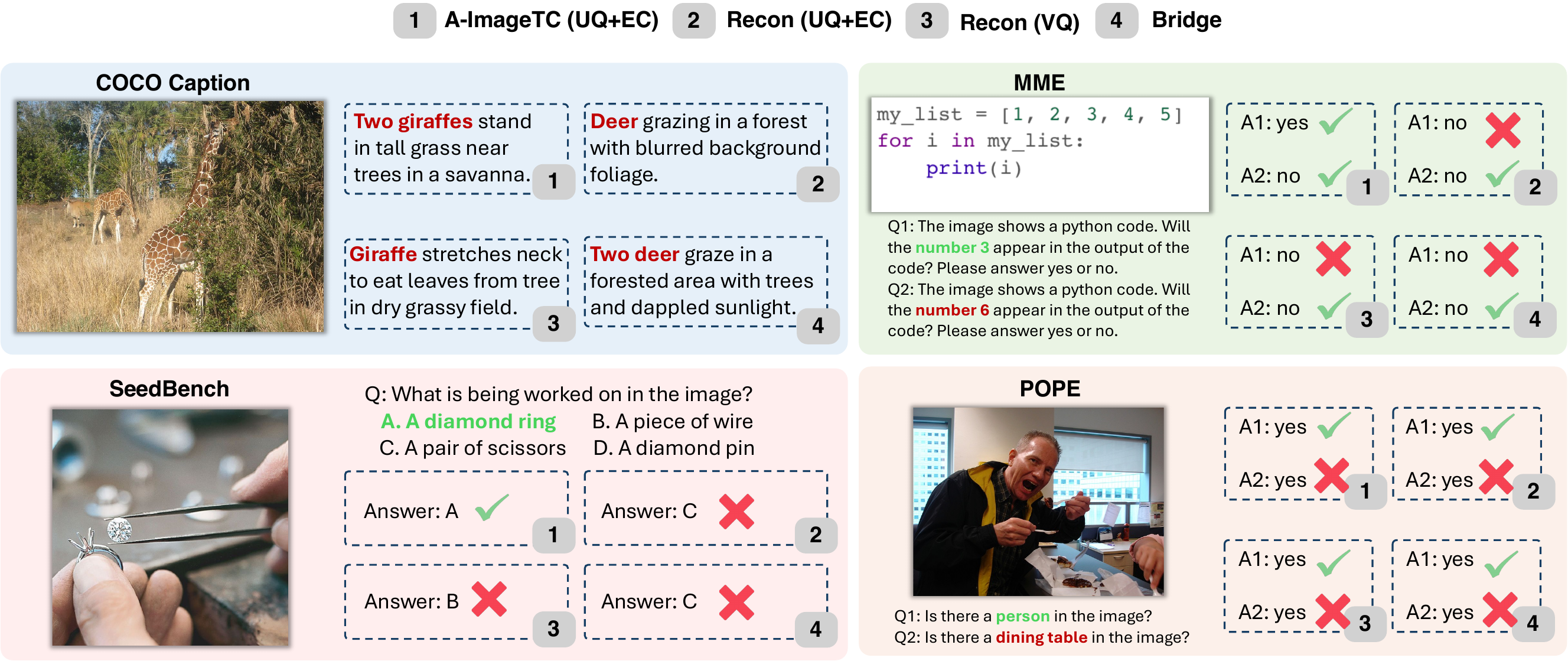}
    \caption{Qualitative comparison of different image processing schemes on COCO Caption, MME, SeedBench, and POPE. Both successful and failed cases are presented.}
    \label{fig:performance_visual}
\end{figure*}

The qualitative examples are shown in Fig. \ref{fig:performance_visual}.
For the COCO Caption task, the proposed method correctly captures the main object \textit{giraffe}, and also describes its number accurately.
For the code understanding example in MME, the proposed method also gives the correct answers.
By contrast, the VQ-based reconstruction scheme has difficulty reconstructing textual images, which makes it difficult to complete the MME task.

For POPE, we further present a failure case where all methods give the wrong answer.
For the dominant object \textit{person}, all methods correctly answer Q1.
However, the eating action of the person misleads all methods and causes hallucination about the existence of a dining table.
This example shows that, although A-ImageTC can better preserve task-relevant visual semantics in many cases, object hallucination under ambiguous visual contexts remains challenging.

\begin{table}[t]
\centering
\caption{Ablation results measured by BD-rate.}
\label{tab:BD-rate}
\resizebox{\linewidth}{!}{
\begin{tabular}{lcccc}
\toprule
\multirow{2}{*}{\textbf{Model variant}}
& \multicolumn{4}{c}{\textbf{BD-rate (\%) $\downarrow$}} \\
\cmidrule(lr){2-5}
& \textbf{MME} & \textbf{POPE} & \textbf{SeedBench} & \textbf{COCO Caption} \\
\midrule
Recon (ELIC)
& 0.00 & 0.00 & 0.00 & 0.00 \\
A-ImageTC
& \textbf{-45.33} & \textbf{-31.91} & \underline{-12.42} & \underline{-46.75} \\
A-ImageTC (w/o SigLIP)
& - & - & - & - \\
A-ImageTC (only SigLIP)
& \underline{-39.58} & \underline{-30.10} & \textbf{-15.00} & \textbf{-49.56} \\
A-ImageTC (CE)
& -31.07 & -24.12 & -7.19 & -17.31 \\
A-ImageTC (w/o Adapter)
& 0.18 & -0.62 & 0.00 & 1.09 \\
A-ImageTC (w/o Recon)
& 7.61 & 10.89 & 60.50 & 5.08 \\
\bottomrule
\end{tabular}
}
\end{table}

\subsection{Ablation Studies and Complexity Analysis}
We conduct ablation experiments on both the architecture and the loss design of the proposed framework.
Recon (ELIC) is selected as the anchor, and the Bjøntegaard delta rate (BD-rate) is calculated for each variant.
As shown in Table \ref{tab:BD-rate}, the model cannot converge properly when the SigLIP loss is removed.
This result indicates that semantic alignment is essential for adapting compression-oriented latents to MLLM-oriented visual tokens.
Replacing the proposed SigLIP-based alignment with the CLIP-style  CE alternative leads to clear performance degradation on every benchmark.
This demonstrates the effectiveness of the proposed semantic alignment loss.
For the architectural ablation, removing either the adapter or the reconstruction prior significantly weakens the model.

\begin{table}[t]
\centering
\caption{Complexity and runtime comparison.}
\label{tab:complexity}
\resizebox{\linewidth}{!}{
\begin{tabular}{lcccc}
\toprule
\multirow{2}{*}{\textbf{Model variant}}
& \multicolumn{2}{c}{\textbf{Full Pipeline}}
& \multicolumn{2}{c}{\textbf{Adapter}} \\
\cmidrule(lr){2-3}
\cmidrule(lr){4-5}
& \textbf{Tx Time (ms)}
& \textbf{Rx Time (ms)}
& \textbf{MACs (M/pixel)}
& \textbf{Params (M)} \\
\midrule
A-ImageTC (DCVC-RT)
& 20.12 & 194.78 & 0.063 & 18.19 \\
A-ImageTC (adaptive)
& 20.57 & 190.93 & 0.063 & 20.85 \\
Recon (DCVC-RT)
& 20.18 & 193.05 & -- & -- \\
Uncompressed
& -- & 170.25 & -- & -- \\
\bottomrule
\end{tabular}
}
\end{table}

We further evaluate the model complexity and runtime.
The resolution of the test image is $256\times256$, and the evalutation is conducted on an NVIDIA A100 80GB GPU.
The results are compared with the existing MLLM interaction pipeline based on reconstruction, as shown in Table \ref{tab:complexity}.
It can be observed that the adaptive adapter introduces about $2$M additional parameters.
However, the computational complexity remains almost unchanged.
Compared with Recon, adding the adapter has little influence on the receiver-side processing time.
These results indicate that the proposed adapter-based token injection brings only a small computational and storage overhead in the full MLLM interaction pipeline.

\section{Conclusion}
\label{sec:Conclusion}

In this paper, we investigate MLLM-oriented token communication to reduce the transmitted data required for MLLM interaction.
By integrating a neural codec into the vision tokenizer, the number of transmitted bits is reduced.
At the receiver, instead of only reconstructing a compressed image, the proposed framework uses decoded latents to construct visual tokens and merge them with reconstruction priors.
To better fuse visual tokens with the reconstruction prior and provide suitable visual inputs for the MLLM, a two-stage visual-language alignment scheme is proposed.
The distillation stage stabilizes injected tokens by matching intermediate features of the original-image branch.
The semantic alignment stage aligns final visual semantics with textual semantics through a SigLIP-based loss.
To adapt to various codec rates, a QP-adaptive adapter based on FiLM is also designed.
Simulation results on multiple MLLM benchmarks show that the proposed method outperforms reconstruction-based coding and other baselines under the same amount of transmitted data.

\bibliographystyle{IEEEtran}
\bibliography{reference}

\end{document}